\documentclass{aa}

\usepackage{graphicx}
\usepackage{txfonts}
\usepackage{booktabs}
\usepackage{xcolor}
\usepackage{lastpage}
\usepackage[
    colorlinks=true,
    linkcolor=blue,
    citecolor=blue,
    filecolor=blue,
    urlcolor=blue
    ]{hyperref}
\usepackage{graphicx}
\usepackage{amsmath}
\usepackage{orcidlink}
\usepackage{multicol}
\usepackage{enumitem}
\usepackage{subcaption}
\newcommand{\msun}{M$_{\odot}$}
\newcommand{\msunf}{\text{M}_{\odot}}

\newcommand{\zmw}{$\langle \mathrm{[Z/H]} \rangle_{\mathrm{MW}}$}
\newcommand{\zlw}{$\langle \mathrm{[Z/H]} \rangle_{\mathrm{LW}}$}

\newcommand{\UCM}{\label{UCM} Departamento de F\'{\i}sica de la Tierra y Astrof\'{\i}sica, Universidad Complutense de Madrid, E-28040, Spain}
\newcommand{\IPARC}{\label{IPARC} Instituto de F\'{\i}sica de Part\'{\i}culas y del Cosmos, IPARCOS,  E-28040, Spain}
\newcommand{\ITA}{\label{ITA} Universit\"{a}t Heidelberg, Zentrum f\"{u}r Astronomie, Institut f\"{u}r Theoretische Astrophysik, Albert-Ueberle-Str.\ 2, 69120 Heidelberg, Germany}
\newcommand{\UW}{\label{UW} Department of Physics and Astronomy, University of Wyoming, Laramie, WY 82071, USA}
\newcommand{\IWR}{\label{IWR} Universit\"{a}t Heidelberg, Interdisziplin\"{a}res Zentrum f\"{u}r Wissenschaftliches Rechnen, Im Neuenheimer Feld 225, 69120 Heidelberg, Germany}
\newcommand{\IAC}{\label{IAC} Instituto de Astrof\'isica de Canarias, calle Vía L\'actea s/n, E-38205 La Laguna, Tenerife, Spain}
\newcommand{\ULL}{\label{ULL} Departamento de Astrof\'isica, Universidad de La Laguna, Avenida Astrof\'isico Francisco S\'anchez s/n, E-38206 La Laguna, Spain}
\newcommand{\Ox}{\label{Ox} Sub-department of Astrophysics, Department of Physics, University of Oxford, Keble Road, Oxford OX1 3RH, UK}
\newcommand{\UniCA}{\label{UniCA} Université Côte d'Azur, Observatoire de la Côte d'Azur, CNRS, Laboratoire Lagrange, France}
\newcommand{\TKU}{\label{TKU} Department of Physics, Tamkang University, No.151, Yingzhuan Road, Tamsui District, New Taipei City 251301, Taiwan}
\newcommand{\INAF}{\label{INAF} INAF–Osservatorio Astrofisico di Arcetri, Largo E. Fermi 5, I-50157, Firenze, Italy}

\begin{document}

\title{Determining star formation histories and age-metallicity relations with convolutional neural networks}


\author{Enrique Galceran \inst{\ref{UCM},\ref{IPARC}} \orcidlink{0000-0002-6550-4539}\thanks{\email{egalcera@ucm.es}}
       \and Patricia S\'anchez-Bl\'azquez\inst{\ref{UCM},\ref{IPARC}}\orcidlink{0000-0003-0651-0098}
       \and Artemi Camps-Fari\~na\inst{\ref{UCM},\ref{IPARC}}\orcidlink{}
       \and Médéric Boquien \inst{\ref{UniCA}} \orcidlink{0000-0003-0946-6176}
       \and Ralf S.\ Klessen \inst{\ref{ITA},\ref{IWR}} \orcidlink{0000-0002-0560-3172}
       \and Francesco Belfiore \inst{\ref{INAF}} \orcidlink{0000-0002-2545-5752}
       \and Daniel~A. Dale \inst{\ref{UW}} \orcidlink{0000-0002-5782-9093}    
       \and Francesca Pinna \inst{\ref{IAC}, \ref{ULL}} \orcidlink{0000-0001-5965-3530}
       \and Ivan S. Gerasimov \inst{\ref{UniCA}} \orcidlink{0000-0001-7113-8152}
       \and Thomas G. Williams\inst{\ref{Ox}}\orcidlink{0000-0002-0012-2142}
       \and Hsi-An Pan \inst{\ref{TKU}} \orcidlink{0000-0002-1370-6964}
}

\institute{\tiny
\UCM     \and  
\IPARC  \and
\UniCA \and
\ITA \and
\IWR \and
\INAF \and
\UW \and
\IAC \and
\ULL \and
\Ox \and
\TKU
}

    \titlerunning{Determining Star Formation Histories And Metallicity With CNN}

   \date{Received Decembre 4, 2025; accepted May 6, 2026}

 
\abstract
   {Recovering the star formation and chemical enrichment histories of galaxies is essential for understanding the physical processes that govern their formation and evolution. Classical full spectral fitting techniques have enabled major advances in this field, but the inversion problem remains highly degenerate, particularly when the available data have limited wavelength coverage and moderate signal-to-noise ratios.}
   {We aim to develop a state-of-the-art tool to infer detailed star formation histories (SFHs) and age-metallicity relations from realistic observational data, while mitigating classical degeneracies and substantially reducing computational cost. In particular, we seek to exploit the complementarity of spectroscopic and photometric data to improve constraints on the spatially resolved SFH and metallicity evolution of nearby galaxies in the PHANGS collaboration.}
   {We construct and train a convolutional neural network (CNN) that combines convolutional layers, attention mechanisms, and a shared latent space to jointly predict SFHs and metallicities in 16 age bins. The network simultaneously processes integral-field spectroscopic data from PHANGS-MUSE and five-band photometric fluxes from PHANGS-HST. Training is performed on a dataset of 165\,000 synthetic spectra and photometric measurements spanning a broad range of SFH shapes, metallicity evolution, dust attenuation, and signal-to-noise ratios representative of the observations.}
   {The CNN accurately recovers SFHs and age-metallicity relations over a wide range of evolutionary scenarios. The inferred luminosity- and mass-weighted mean ages and metallicities show negligible bias, with dispersions of $\sim0.12$ dex in age and $\sim0.03$ dex in metallicity. When applied to real PHANGS-MUSE and PHANGS-HST data for NGC\,3627, the network produces smooth, spatially coherent maps of stellar age and metallicity that recover physically meaningful structures, including younger populations tracing the spiral arms and star-forming regions. The CNN is approximately $5\times10^{3}$--$2\times10^{4}$ times faster than traditional full spectral fitting codes, providing a powerful and efficient alternative for the analysis of large spectro-photometric surveys.}
   {}


\maketitle
%

\section{Introduction}
The study of the star formation histories (SFHs) of galaxies offers crucial insights into their evolutionary pathways by enabling us to reconstruct how they assembled, while also highlighting the role of diverse physical processes in shaping the properties observed today \citep[see e.g.][]{CidFernandes2004, Hoopkins2006, Asari2007, Panter2003, SB2011, Carnall2018, Pessa2023}. 

In the Milky Way and nearby Local Group galaxies, SFHs can be reconstructed in detail by fitting isochrones to colour–magnitude diagrams (CMDs) of resolved stars \citep[see review by][]{Tolstoy2009}. For the vast majority of galaxies, however, the individual stars cannot be resolved, and SFHs must instead be inferred by comparing their spectral energy distribution (SED) or the integrated spectral information with predictions from stellar population models. 

Establishing a connection between models and observations requires a statistical inference framework. Full spectral fitting has become a widely used approach for deriving stellar population properties from integrated spectra, typically by comparing observational data with a combination of Single Stellar Population (SSP) models \citep[e.g.][]{Ocvirk2006a, Ocvirk2006b, Tojeiro2007, Wilkinson2017, Cappellari2023, CidFernandes2023}. However, the inference of SFH from integrated spectra is an ill-conditioned problem, where small uncertainties or noise in the input spectrum can lead to large, unstable, or non-unique variations in the derived SFH \citep[e.g.][]{Ocvirk2006a, Ocvirk2006b}. Some codes alleviate this problem by establishing a regularisation of the inversion, favouring smoothly varying solutions over other equally probable solutions \citep[e.g.][]{Ocvirk2006b, Cappellari2023}; however, their success depends heavily on the signal-to-noise ratio of the input data. More recently, stellar population inference methods have increasingly adopted Bayesian formalisms \citep[e.g.][]{Turner2021, Johnson2021, Maksymowicz-Maciata2024, Iglesias-Navarro2024}, where Markov Chain Monte Carlo (MCMC) sampling is commonly employed to characterise posterior distributions. Under the assumption of approximate Gaussianity, this allows for an efficient mapping of the degeneracies inherent in the high-dimensional parameter space, but it is very time-consuming.

Challenges persist, and different codes usually recover different solutions depending on the strategies used to fit the data, even when using the same model templates \citep[see e.g.][]{RuizLara2015, Sanchez2016RM, Magris2015, Ge2018}. Certainly, the problems arise primarily from the existing degeneracies between the parameters defining the solution \citep{Worthey1992} and the limited ability of fitting strategies to distinguish subtle spectral differences or differences spread across correlated pixels.

Machine Learning (ML) techniques have seen significant advances, propelling them into everyday use. Scientific applications have also benefited from this improvement, which automates procedures that have traditionally required human supervision and improves the accuracy and speed of data analysis.  Convolutional Neural Networks (CNN), due to their capacity for feature extraction and pattern recognition, have proven to be versatile and extremely successful for several astrophysical tasks, such as morphological classification of galaxies \citep[][]{Dieleman2015, Huertas-Company2015, Cabayol2019, Cheng2020, Vega-Ferrero2021, Walmsley2022, DominguezSanchez2022, Hausen2020}; detection of low surface brightness features \citep{DominguezSanchez2023}; stellar cluster classification \citep{Hannon2023}; identification of deeply obscured clusters \citep{graham2025}; determination of stellar parameters \citep{Fabbro2018}; and stellar cluster identification \citep{Maschmann2024}, among others. 

The application of CNNs to stellar population inference has also shown promising results. Several studies have demonstrated that stellar ages and metallicities can be recovered with an accuracy comparable to, or even exceeding, that of traditional full spectral fitting methods when using photometric data as input \citep[see, e.g.,][]{Liew-Cain2021, Walter2026}. More recently, \citet{Woo2024} reported substantial improvements in both the bias and the scatter of mean age and metallicity estimates derived from spectra relative to conventional fitting techniques.
The work of \citet{Lovell2019} pioneered the use of CNNs to recover not only average stellar ages and metallicities, but also the full star formation history. However, their model assumed a single, constant metallicity.  To the best of our knowledge, the ability of CNNs to infer the same level of detail as full spectral fitting methods, i.e,  both the full SFH and the age-metallicity relation, has so far been explored only in \citep{Anwar2025}.

In this work, we present a new CNNs to infer SFH and age-metallicity relations from a combination of spectra observed with the MUSE integral field spectrograph at the ESO VLT and HST photometric fluxes in 5 NUV/optical bands that mimic the data taken as part of the  Physics at High Angular Resolution in Nearby Galaxies Surveys\footnote{\url{http://www.phangs.org}} (PHANGS).

The PHANGS survey aims to provide a unified picture of star formation across the different phases of the interstellar medium in a representative sample of nearby, massive, star-forming galaxies \citep[see]{Leroy2021}. Its main objectives are to constrain star-formation timescales and efficiencies, assess the role of stellar feedback processes, trace chemical enrichment and mixing across galaxy discs, and understand how young stellar clustering both emerges from and reshapes the surrounding medium. To this aim, the survey is obtaining multi-wavelength observations able to resolve key structures such as molecular clouds, H II regions, and star clusters on scales of a few to $\sim$100 pc in a sample of star-forming galaxies\citep[see survey papers][]{Leroy2021, Lee2022, Lee2023, Emsellem2022, Chandar2025}. It also samples a wide range of galactic environments---bars, spiral arms, and centres---to evaluate environmental effects, while ensuring consistency with the broader population of star-forming galaxies along the main sequence  Our work directly supports several key goals of the PHANGS project, including constraining the timescales of star-formation processes, quantifying the role of different stellar feedback mechanisms, and resolving chemical enrichment and mixing across galaxy discs.

This article is structured as follows. Section~\ref{sec:data} describes the observational datasets that motivate this work, together with the construction and encoding of the synthetic training set designed to reproduce their spectro-photometric characteristics. Section~\ref{sec:model} details the architecture and implementation of the CNN and outlines the configuration of the {\tt pPXF} full spectral fitting used for comparison. The performance of both approaches is evaluated in Sec.~\ref{sec:Results}, where we also explore the robustness of the CNN to different input conditions. In Sec.~\ref{sec:NGC3627} we compare the performance of our trained CNN using real observational data from the nearby spiral galaxy NGC\,3627. Finally, Sec.~\ref{sec:Discussion} discusses the advantages and limitations of our method and its prospects for deriving realistic SFHs and age-metallicity relations. In this paper, we assume a flat $\Lambda$CDM cosmology with $\Omega_M$ = 0.3, $\Omega_{\Lambda}$ = 0.7, and a Hubble constant H$_0$ = 70 km s$^{-1}$Mpc$^{-1}$.

\section{Data}\label{sec:data}

 In this work, we simulate observations designed to reproduce the characteristics of the PHANGS-MUSE (\citealt{Emsellem2022}, hereafter Em22) and PHANGS-HST \citep{Lee2022} surveys.

The PHANGS-MUSE dataset comprises 168 pointings obtained with the MUSE integral-field spectrograph at the ESO/VLT, targeting 19 nearby galaxies in the redshift range $0.0012 \leq z \leq 0.0046$ and spanning stellar masses $9.4 < \log(M_{\star}/M_{\odot}) < 11.0$. These observations achieve a median spatial resolution of $\sim50$ pc. The spectra cover the wavelength range 4750–9350~\AA{}, with a spectral resolution varying from FWHM $\sim2.9$~\AA{} at 4800~\AA{} to $\sim2.4$~\AA{} at $\lambda > 6500$~\AA{}. In this work, we restrict our analysis to $\lambda < 7000$~\AA{} in order to minimise contamination from sky-subtraction residuals (see Em22).

The PHANGS-HST survey provides $NUV$–$U$–$B$–$V$–$I$ imaging for 38 nearby spiral galaxies, including the full PHANGS-MUSE sample, with a median spatial resolution of $\sim5$ pc. The inclusion of the NUV band is particularly valuable, as emission in this wavelength range is highly sensitive to small fractions of young stellar populations, making it a powerful tracer of recent, low-level star formation that may remain undetected in optical spectra \citep[e.g.][]{Werle2019, SalvadorRusinol2020}. Moreover, while different combinations of age and metallicity can produce nearly indistinguishable optical spectra, they give rise to significantly different spectral energy distributions in the NUV. Consequently, the inclusion of NUV photometry helps to break the age–metallicity degeneracy and improves the robustness of the inferred stellar population properties \citep[e.g.][]{Kaviraj2007, Angthopo2024}.

\subsection{Generation of synthetic SFH}
\label{sec:GenerationOfSyntheticSFH}
Supervised Machine Learning requires a training set with known inputs (the observable quantities) and corresponding outputs (labels). In our case, the labels are the stellar mass fractions and mean metallicities in 16 age bins. To construct this dataset, we generated a diverse set of synthetic SFHs and computed the associated spectra and HST fluxes under different assumptions of chemical enrichment and dust attenuation.

We employed the {\tt GP-SFH} module of {\tt DenseBasis} \citep{Iyer2019}, which parametrises SFHs using five quantities: the total stellar mass, the specific star formation rate (sSFR) at z=0, and three look-back times at which the cumulative stellar mass reaches fixed fractions of the final mass. We fixed these fractions to 1/4, 1/2, and 3/4.

Since our objective is not to infer the total stellar mass\footnote{The stellar mass can be determined a posteriori from the inferred mass-to-light ratio and the observed flux.}, we set this parameter to $15\cdot 10^6\msunf$. The remaining parameters were chosen randomly from the following distributions.  

The three look-back times were drawn from a Dirichlet distribution \citep[Multivariate Beta;][]{Kotz2000} with concentration parameter $\alpha=1$. Although \citet{Iyer2019} recommend $\alpha = 5$ for SED fitting applications, we found that $\alpha=1$  provides a more balanced mixture of narrow and extended SFH profiles for the purpose of generating a wide variety of training examples.

The sSFR at z = 0  was sampled from a uniform distribution in the range $10^{-12}~\mathrm{yr}^{-1} \leq \mathrm{sSFR} \leq 10^{-9}~\mathrm{yr}^{-1}$. Although galaxies in the PHANGS sample are selected to lie on the star-forming main sequence (with global sSFR typically above $10^{-11}~\mathrm{yr}^{-1}$), our analysis focuses on spatially resolved regions within galaxies,  where local sSFR values can vary above and below the global average.

Using {\tt GP-SFH}, we generated 150\,000 synthetic galaxies. However, the resulting colour distribution did not sufficiently populate the reddest and bluest regions of the PHANGS colour space. To better sample these extremes, we generated an additional 15\,000 SFHs using skew-normal functions, designed to represent regions with very old populations and red colours or regions where young stellar clusters dominate the light.

Skew-normal functions are characterised by a peak age, a width, and a skewness parameter, allowing for flexible control of both the timing and asymmetry of the SFH. The peak age was drawn from a Gaussian distribution with mean $\mu = 12$~Gyr and standard deviation $\sigma = 3$~Gyr, ensuring that the bulk of the star formation occurs at early cosmic times. The width of the distribution was also sampled from a Gaussian distribution, with $\mu = 1$~Gyr and $\sigma = 0.4$~Gyr, favouring relatively short formation timescales. The skewness parameter was drawn from a uniform distribution in the range $-0.25 \leq \alpha \leq 0.75$. This asymmetric range was intentionally chosen to preferentially generate SFHs skewed towards earlier times, thereby enhancing the number of systems dominated by old stellar populations.

In 40\% of these SFHs, we superimposed a secondary, short-duration burst of recent star formation (40--50~Myr) occurring within the last 500~Myr. The burst age was sampled from a log-uniform distribution, while its width and mass fraction (ranging from 1--10\% of the total stellar mass) were drawn from uniform distributions within the specified limits. All parameter distributions are summarised in Table~\ref{Table:parameters_and_distributions}.

In total, we generated 165,000 SFHs:
\begin{itemize}
\item 150\,000 with {\tt GP-SFH},
\item 9\,000 with the skewed-normal parametrisation, and
\item 6\,000 with a skewed-normal SFH plus an added burst.
\end{itemize}

All SFHs were normalised to the same total mass as \texttt{DenseBasis}. The resulting training set spans a wide variety of evolutionary shapes---from smooth, gradually declining histories to rapidly varying cases with young bursts at different epochs. We did not constrain the SFHs to follow physically motivated evolutionary tracks, as our primary goal is to train the network to recognise how variations in the SFH at different times imprint themselves on the spectrum and photometry. Furthermore, the network is designed to analyse spatially resolved SFHs and age-metallicity relations, whose shapes can differ significantly from galaxy-integrated averages.

\begin{table}[htbp]
\caption{\label{Table:parameters_and_distributions} Distributions of parameters for skewed-normal SFHs.}
\centering
\begin{tabular}{lccc}
\toprule\toprule
Parameter & Distribution & $\mu$ / min & $\sigma$ / max   \\
\midrule
\multicolumn{4}{l}{Skewed–normal SFH parameters}\\
\midrule
Peak age   & normal  & 12 Gyr   & 3 Gyr    \\
Width      & normal  & 1 Gyr    & 0.4 Gyr   \\
Skewness   & uniform & $-0.25$  & 0.75      \\
\midrule
\multicolumn{4}{l}{Burst parameters}\\
\midrule
Age           & log-uniform & 10 Myr   & 500 Myr \\
Width          & uniform     & 40 Myr   & 50 Myr   \\
Mass fraction  & uniform     & 1\%      & 10\%    \\
\bottomrule\bottomrule
\end{tabular}
\tablefoot{For Gaussian distributions, $\mu$ and $\sigma$ denote the mean and standard deviation; for uniform and log-uniform distributions, the listed values indicate the minimum and maximum limits.}
\end{table}


\subsection{Generation of synthetic observations}\label{sec:GenerationOfSyntheticObservations}

For each SFH, we used {\tt ProSpect} \citep{Robotham2020} to generate the synthetic spectra and photometric fluxes used as inputs to our CNN. {\tt ProSpect} computes synthetic spectra from specified star formation and metallicity histories, incorporating both dust attenuation and emission through the \citet{Charlot2000} two-component attenuation model and the \citet{Dale2014} templates, respectively, as well as ionised gas emission lines using the {\tt MAPPINGS-III} \citep{Allen_2008} photoionisation tables. Other widely used codes, such as {\tt MAGPHYS} and {\tt CIGALE} \citep{daCunha2008, Noll2009, Boquien2019}, employ similar strategies to produce synthetic observables. However, {\tt ProSpect} provides great flexibility to customise and extend the available templates for stellar and dust emission, and to implement a wide range of prescriptions for SFH and age-metallicity relations.

 We use the {\tt E-MILES} stellar population models \citep{Vazdekis2010, Vazdekis2016}, with the extension to young ages (6.3 Myr to 63 Myr) as described in \cite{Asad2017}. The {\tt E-MILES} templates cover the wavelength range $\lambda = 1680-50\,000$~\AA{} with a spectral resolution of FWHM = 2.5\AA{}, combining four empirical stellar libraries: NGSL \citep{Gregg2006}, MILES \citep{Sanchez-Blazquez2006}, Indo-US \citep{Valdes2004}, and CaT \citep{Cenarro2001}. We adopted the version based on the Padova2000 isochrones \citep{Girardi2000}, which spans ages from 63 Myr to 17.8 Gyr and includes six metallicities. The reason is that this version can be reliably joined at 63 Myr with the young extension (ages between 6.3 and 63 Myr; \citet{Asad2017}), which uses scaled-solar isochrones from \cite{Bertelli1994}, extended to low-mass stars (M < 0.5\msun) with the stellar tracks from \cite{Pols1995}. We use the versions computed with a \cite{Chabrier2001} initial mass function (IMF).

Dust extinction is incorporated by multiplying each model spectrum by the transmission function
\begin{equation}
T(\lambda) = \exp\left[-\tau_{\lambda_{\mathrm{piv}}}\left(\frac{\lambda}{\lambda_{\mathrm{piv}}}\right)^{n}\right],
\end{equation}
where $\tau_{\lambda_{\mathrm{piv}}}$ is the optical depth at $\lambda_{\mathrm{piv}} = 5500,\text{\AA}$  and $n$ is the attenuation slope. We adopt the two-component attenuation prescription of \cite{Charlot2000}, where the optical depth includes a diffuse component, $\tau_{\rm ISM}$, and, for stellar populations younger than 10 Myr, an additional birth-cloud contribution, $\tau_{\rm BC}$. We fix  $n=0.7$, consistent with the empirical correlation between the far-infrared–to–ultraviolet luminosity ratio and the UV spectral slope in star-forming galaxies \citep{Charlot2000}. 

\begin{figure}
\centering
\includegraphics[width=0.45\textwidth]{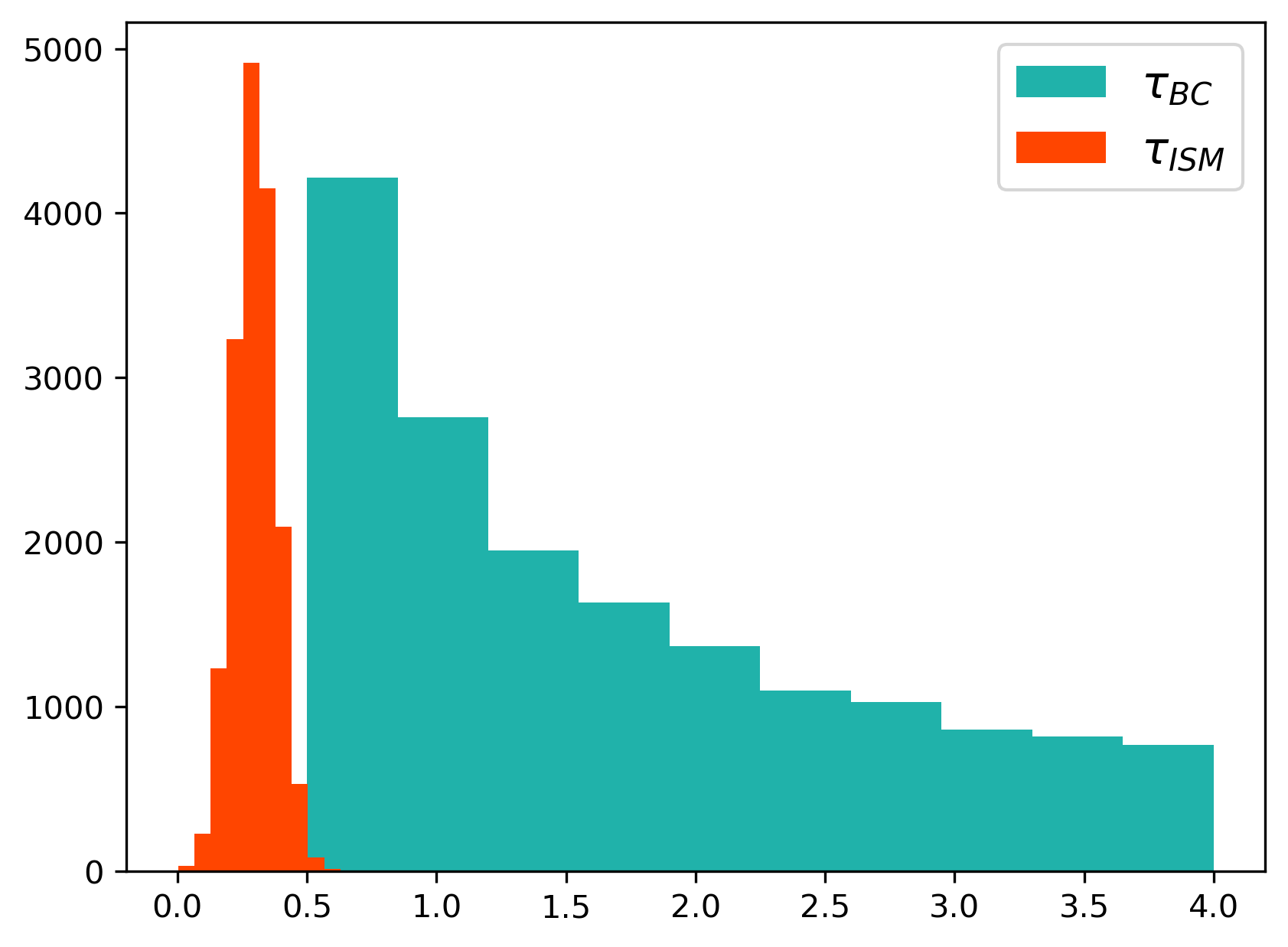}
\caption{Distribution of optical depths for the diffuse and birth-cloud components of the \citet{Charlot2000} attenuation curve used to simulate internal extinction in the synthetic data.}
\label{fig:tau_distribution}
\end{figure}

The fiducial model of \citet{Charlot2000} adopts $\tau_{\rm ISM} = 0.3$ and $\tau_{\rm BC} = 1$, values that reproduce the typical attenuation of O-type and supergiant stars in the Milky Way. However, the effective optical depth is expected to vary from galaxy to galaxy, depending on global properties such as metallicity and gas content, as well as geometric factors like inclination. To account for this diversity, we introduce object-to-object variations in the synthetic dataset by sampling these parameters from physically motivated distributions. Specifically, $\tau_{\rm BC}$ is drawn from a log-uniform distribution in the range $0.5$-$4.0$, while $\tau_{\rm ISM}$ is sampled from a Gaussian distribution centred at $0.3$ with a standard deviation of $0.08$. These ranges are chosen to reproduce the attenuation properties observed in the PHANGS--MUSE sample (see \citealt{Emsellem2022}), inferred from the Balmer decrement for $\tau_{\rm BC}$ and from the stellar $E(B-V)$\footnote{A selective extinction $R_V = 3.1$ is assumed.} derived through full spectral fitting for $\tau_{\rm ISM}$. The resulting distributions are shown in Fig.~\ref{fig:tau_distribution}.

Although the wavelength coverage of our current dataset does not reach the far-infrared (FIR), we account for dust re-emission of absorbed light by incorporating the FIR template library of \citet{Dale2014}. This inclusion enables future extensions of the model to incorporate PHANGS--JWST \citep{Lee2023} observations. At each stage of the attenuation process, we enforce energy conservation by re-emitting the total absorbed luminosity of the stellar radiation field, which determines the temperature of the emitting dust. This allows for different values of $\alpha$ to be applied to the diffuse ISM and the birth cloud.

\texttt{ProSpect} incorporates a simple energy balance scheme to produce nebular emission features for a range of gas-phase metallicities. The key default assumption is that flux short of the Lyman limit is absorbed with an efficiency determined by the UV photon escape fraction. The integrated intrinsic stellar flux is then re-emitted using line energies determined by \texttt{MAPPINGS-III} as per the tables provided by \citet{Levesque2010}.

Rather than generating models with constant metallicities, we implemented a simple prescription in which metallicity increases in tandem with star formation. To avoid enforcing a specific (and potentially unrealistic) enrichment law, we allowed the metallicity yield and the final metallicity to vary among models.

The metallicity evolves with stellar age according to
\begin{equation}
Z(\text{age}) = Z_{\text{start}} + (Z_{\text{final}} - Z_{\text{start}}) \times \frac{m_{\text{age}}}{M_{\text{tot}}},
\label{eq:Z.evolution}
\end{equation}
where $Z_{\text{start}}$ and $Z_{\text{final}}$ represent the metallicities of the oldest and youngest stellar populations, respectively, and $\frac{m_{\text{age}}}{M_{\text{tot}}}$ is the cumulative mass fraction formed up to a given age bin. We fixed $Z_{\text{start}} = 0.008$, while $Z_{\text{final}}$ was randomly sampled between 0.010 and 0.030, effectively allowing for a range of enrichment slopes.

To obtain the HST fluxes, we convolve the synthetic spectra with the transmission curves of the filters available for the PHANGS-HST sample (Wide Field Camera 3; F275W, F336W, F438W, F555W, and F814W).

\subsection{Preprocessing of synthetic data}
To mimic MUSE observations,  we resampled the synthetic spectra to match the wavelength-dependent resolution and sampling of the MUSE observations (4800\AA{} to 7000\AA{} with a step of 1.25\AA{}). 
Furthermore, we applied an additional broadening to all synthetic spectra up to a common velocity dispersion of 150 km~s$^{-1}$. We did not add any velocity shift to the spectra. 

These same preprocessing steps must be applied to the observed spectra before using the CNN to infer their properties. Specifically, the spectra need to be shifted to the rest frame, broadened to a common velocity dispersion of $\sigma_{\rm f}$ = 150 km~s$^{-1}$, using  $\sigma_{\rm broadening} = \sqrt{150^2 - \sigma_{\rm gal}^2}$ (km~s$^{-1}$), were $\sigma_{\rm gal}$ represent the stellar velocity dispersion. 

Additionally, each spectrum was normalised by its mean flux in the 5300--5500\AA{} window, chosen because it aligns with the F555W pivot wavelength and is free of strong emission lines.

As our goal is to infer stellar population properties from real galaxies, we incorporate noise into the synthetic data to better mimic observational conditions. Although the PHANGS-MUSE spectra are Voronoi-binned to achieve a minimum signal-to-noise ratio (SNR) of $\sim35$ per\,\AA, internal tests show that training on a range of SNR values leads to improved model performance compared to training at a single value. We therefore add Poisson noise to the synthetic spectra, scaling its amplitude to produce SNR values of 15, 20, 35, and 50.  In Sec.~\ref{sec:Validation}, we further assess the performance of the model in relation to the noise level of the input data, and to more realistic noise patterns derived from the residuals of fits to real data.

\section{Neural network model}\label{sec:model}

Our CNN  architecture combines convolutional, pooling, attention, and fully connected (dense) layers. A brief overview of their functions is provided below, but we refer to \citet[][]{dumoulin2016guide, khan2019survey, li2020survey, zhao2024review} for a more detailed explanation.

A convolutional layer is the fundamental building block of a CNN. It applies a set of learnable filters (or kernels) that convolve across the input data to extract local features such as edges, gradients, or spatial patterns \citep{dumoulin2016guide}. Each filter produces a feature map that highlights specific structures in the input, enabling the network to learn hierarchical representations: from low-level features in the first layers to increasingly complex patterns in deeper layers \citep{Lecun1998}. This local connectivity and weight sharing make convolutional layers particularly effective for analysing structured data such as images, spectra, or spatial maps. We use one-dimensional (1D) convolutions, which are well-suited to spectral data. 

A max-pooling layer is a downsampling operation commonly used in CNNs to reduce the spatial dimensionality of feature maps while retaining the most relevant information. It divides the input into small regions and outputs the maximum value within each region, effectively preserving the strongest responses of the convolutional filters. This operation allows our network to focus on spectral or spatial features that exhibit the maximum response to variations in stellar population properties, while enhancing translational invariance, reducing computational cost, and mitigating overfitting \citep{dumoulin2016guide}.

A fully connected (dense) layer connects every neuron in one layer to every neuron in the next, allowing the network to combine and interpret the features extracted by previous convolutional or pooling layers \citep{rumelhart1986learning}. Each neuron computes a weighted sum of its inputs, followed by a nonlinear activation, enabling it to learn complex relationships among features. In this work, fully connected layers integrate spectral patterns identified in earlier stages of the network to produce global predictions, such as SFH and age-metallicity relations.

An attention mechanism (originally developed for sequence processing in natural language tasks for large language models \citep{Vaswani2017}) is a computational module designed to enhance the performance of NNs by allowing them to focus selectively on the most informative parts of the input data. Instead of processing all features with equal weight, the network learns to assign higher importance to specific regions or features that are most relevant for the task at hand. In astrophysical applications, attention mechanisms enable the model to identify and emphasise key spectral or spatial signatures while suppressing less informative regions. Furthermore, attention enhances the network’s ability to capture long-range dependencies---such as correlations between spectral features at widely separated wavelengths---that may be inaccessible to local convolution operations alone \citep{Melchior2023}. These dependencies change dynamically; that is, the network focuses on the relevant features on a case-by-case basis.

For an in-depth analysis of Deep Learning and CNNs, we refer to the excellent review \citet{Lecun2015}.

\subsection{Architecture}\label{sec:Architecture}

Our network takes as input a spectrum and 5 photometric bands (see Sec.~\ref{sec:data}) and gives as output the stellar mass contribution and their average metallicity in 16 different age bins with central values:  0.007, 0.021, 0.0425, 0.076, 0.1145, 0.1615, 0.23, 0.345, 0.545, 0.865, 1.37, 2.17, 3.44, 5.455, 8.645, and 14.2 Gyr. The central ages of the bins were chosen to sample, in approximately uniform intervals, the age range covered by the models on a logarithmic scale. As a result, the bin widths are narrower at younger ages, where the observational characteristics exhibit stronger variations. The total number of bins was set to the maximum value at which the network's results showed no improvement.

We note that the ages of some SSP templates in the model grid formally exceed the age of the Universe. These very old ages should not be interpreted literally, but rather as reflecting the upper limit of the ages permitted by the models. Current uncertainties in the model ingredients can result in offsets of a few gigayears in the age scale; for example, a modest 50--100~K offset in the colour--$T_{\rm eff}$ transformation, well within typical uncertainties, can translate into differences of $\sim$2~Gyr in the inferred ages \citep[see, e.g.,][]{Vazdekis2010}. Although some authors choose to exclude the oldest templates from their analyses, doing so would impose an artificial truncation on the parameter space and introduce a saturation effect, biasing age estimates toward the imposed maximum rather than reflecting the intrinsic uncertainties of the models.

\begin{figure*}[!htbp]
\centering
 \includegraphics[width=17cm]{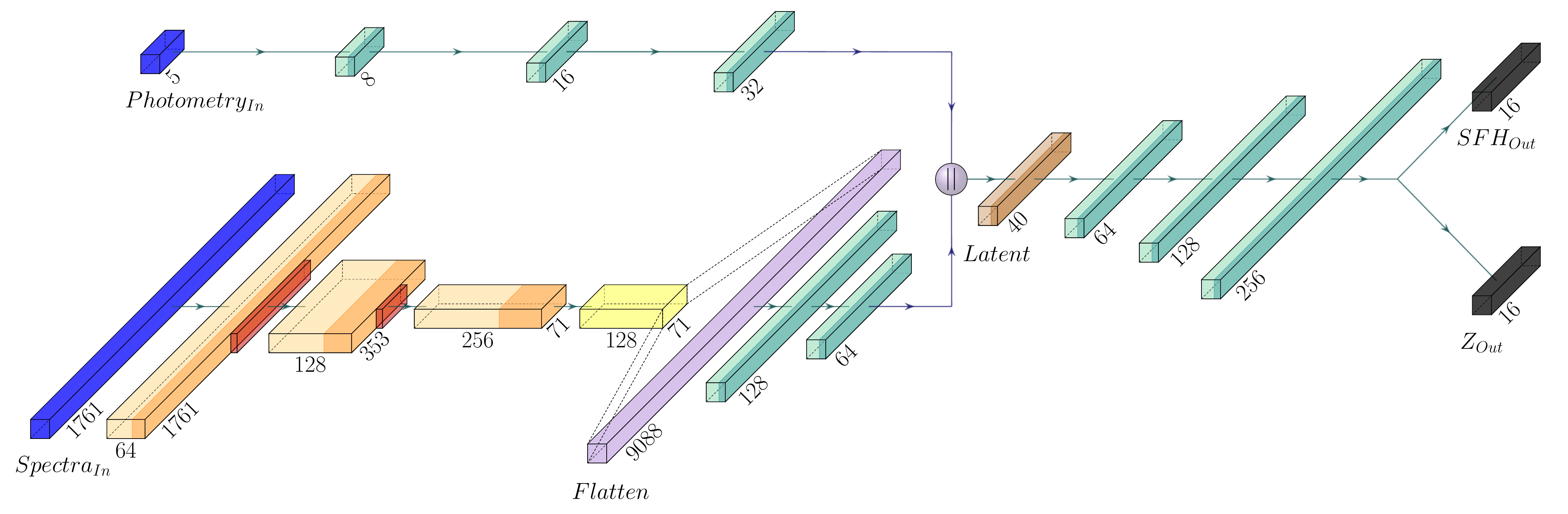}
    \caption{Schematic representation of our CNN Network architecture as described in Section~\ref{sec:Architecture}. The colour code identifies the different types of layers. Orange: convolutional layers, red: max pooling layers, yellow: attention layer, purple: flattening layer, cyan: dense layers. Darker shades, when present, represent the activation and Instance Normalisation. The latent, input and output layers are represented in brown, blue, and black, respectively, while the sphere indicates the position of the concatenation between the spectrum and photometry branches.}
    \label{fig:Red}
\end{figure*}

In Fig.~\ref{fig:Red} we present a schematic representation of the architecture. The model is split into three main sections:
\begin{itemize}
    \item The first section consists of two parallel branches that encode the spectral and photometric information independently. The branch that processes the spectra has three convolutional and max-pooling layers, followed by an attention layer and two fully connected layers. The second branch, dedicated to the photometric data, consists of three fully connected layers. The photometric branch is smaller and shallower, as the number of input data is significantly lower (five photometric inputs vs 1761 spectral points), albeit each element is more information-rich.
    
    \item The middle section of the network contains that latent space, which combines the outputs from the two encoding branches. The features of these outputs (64 features from the spectral branch and 32 from the photometric branch) are concatenated and passed to a single layer referred to as the latent space. In this model, the latent space is a 40-dimensional compressed representation that integrates the most informative features extracted from both inputs. The latent space functions as a bottleneck, constraining the information transmitted to the decoding block to the features most relevant for accurately inferring the target outputs, i.e. the SFH and metallicity evolution.

    \item Finally, the decoding block transforms latent-space information into the target outputs using a single branch composed of three fully connected layers. Combining both outputs within a single branch, rather than defining separate decoders, reduces the number of trainable parameters and mitigates the risk of overfitting. From a physical perspective, this structure also reflects the intrinsic connection between stellar mass growth and chemical enrichment: both the SFH and metallicity evolution are strongly interdependent, and much of the relevant information linking the latent space to the outputs can therefore be shared. Previous studies \citep[e.g.][]{Krizshevsky2012, Hannon2023} have demonstrated that, with appropriate training, multiple correlated quantities can be effectively represented within a shared intermediate layer, without requiring independent processing for each output.
\end{itemize}

Normalising values across layers helps stabilise the learning process, improve training efficiency, and reduce internal covariate shift. Batch Normalisation \citep{Ioffe2015}, which adjusts the distribution of values within each batch, is one of the most widely used methods for this task.  However, we adopted Instance Normalisation \citep[IN,][]{Wu2018}, which normalises each sample independently rather than across a batch (i.e., all samples in a single training step). Note that this normalisation is applied to the features extracted by each layer within the CNN, and it is not related to the normalisation of input data (see Sec.~\ref{sec:GenerationOfSyntheticObservations}). The IN works better than Batch Normalisation when batch statistics vary widely, for example, if the training data includes heterogeneous galaxy spectra. We found that IN provided greater training stability and slightly better predictive accuracy, likely because it reduced inter-sample dependencies and improved the consistency of feature scaling across diverse inputs. Several works have shown that the inclusion of IN often reduces the need for dropout layers \citep[see, e.g.][]{Ioffe2015}. We found that once IN is added, the dropout layers (which set random neurons to zero to reduce overfitting and help with regularisation) did not improve the solution and instead slowed training, leading to their removal. 

The number of layers and neurons per layer (i.e. depth and width of the hidden layers) of each branch were chosen according to the lowest number of each that achieved the best fit. Increasing the number of layers or neurons did not appreciably improve the network's accuracy, so we kept the network as small as possible to avoid overfitting.

\subsection{Activation function}\label{sec:activation}
Activation functions introduce non-linearity into NNs, allowing them to model complex relationships between inputs and outputs. In our model, we employ Parametric Rectified Linear Unit (PReLU; \citet{He2015}), a variant of the standard Rectified Linear Unit (ReLU; $\mathrm{ReLU} \equiv \max(0, x)$) that introduces a learnable parameter controlling the slope for negative input values. While the traditional ReLU sets all negative activations to zero, potentially leading to inactive (`dead') neurons (which leads to the Dying ReLU Problem; \citet{Lu2020}), PReLU allows a small, trainable negative response, enabling the network to retain gradient flow even for negative inputs. The function is defined as:
\begin{equation}
    \mathrm{PReLU}(\alpha, x) = \max(0, x) + \alpha \cdot \min(0, x),
    \label{eq:PReLU}
\end{equation}
where $\alpha$ is a trainable parameter that controls the slope of the negative inputs. There is no clear consensus on what the optimal placement of the activation function is within the network architecture with respect to the intermediate regularising and normalising step (in our case, the IN layer). We explored all possible permutations of the order of the IN, Activation and dropout layers and found differences in loss value (see Sec.~\ref{sec:loss}) to be $\lessapprox5\%$. Ultimately, our final configuration only applies the activation function after the IN Layer to avoid potential artefacts in the normalisation caused by the activation function.

\subsection{Loss function}\label{sec:loss}

The loss function quantifies the discrepancy between the model's predictions and the true target values, providing the basis for optimisation during training. We tested several alternatives and found that the best performance is achieved using the Mean Squared Error (MSE), the Symmetric Mean Absolute Percentage Error (SMAPE), and the Logarithm of the Hyperbolic Cosine (Log-Cosh) loss functions.

MSE is often used as the default loss function for regression tasks because it is simple, differentiable, and easy to optimise with gradient-based algorithms such as gradient descent. However, MSE assumes that the target outputs follow a normal distribution, and as the errors are squared, large deviations have a disproportionately strong impact on the loss. This property can bias the predicted SFH, since larger errors are expected for older stellar populations that are less constrained by the observations. To mitigate this effect, we additionally tested SMAPE and Log-Cosh. The former expresses the deviation between predicted and true values as a percentage, independent of whether the prediction is an over- or underestimation \citep[see, e.g.,][]{Lovell2019}, while the latter provides a smooth approximation to the Mean Absolute Error (MAE) and is less sensitive to outliers than MSE. Both SMAPE and Log-Cosh are therefore considered more robust to outliers and sparsity in the data \citep{Saleh2022}.

No significant differences were found between the predictions of the model trained with SMAPE and Log-Cosh. Still, Log-Cosh had the fastest convergence during the training stage, and we therefore adopted this loss function. Log-Cosh introduces a smoother transition around zero (i.e. perfect prediction), allowing the network to fine-tune its weights more efficiently, thereby reducing the risk of overfitting and improving overall training stability. The Log-Cosh function is defined as:
\begin{equation}
\text{LogCosh}(y, y^p) = \sum_{i=1}^n \log \left[ \cosh \left(y_i^p - y_i\right) \right],
\label{eq:LogCosh}
\end{equation}
where $y$ and $y^p$ are the target and predicted vectors, respectively.
We compute two independent loss values corresponding to the two output vectors---one for the SFH and the other for the age-metallicity relation---and combine them into a single weighted loss function. Since the accurate recovery of the SFH is our primary objective, its contribution to the total loss is increased as much as possible without degrading the metallicity predictions. We found an optimal weighting ratio of 80:20 between the SFH and the age-metallicity relation, yielding a final loss function of the form:
\begin{equation}
\begin{split}
\mathcal{L}\left(y_{\text{SFH}}, y_{\text{Z}}, y^p_{\text{SFH}}, y^p_{\text{Z}}\right) =\ &0.8\cdot \text{LogCosh}\left(y_{\text{SFH}}, y_{\text{SFH}}^p\right) \\
&+ 0.2\cdot \text{LogCosh}\left(y_{\text{Z}}, y_{\text{Z}}^p\right).
\end{split}
\label{eq:LossFunction}
\end{equation}

\subsection{Training}\label{sec:Training}

The network weights were initialised using the uniform scheme of \citet{Glorot2010} and optimised via Adam \citep{Kingma2014} with an initial learning rate of $10^{-4}$ and a linear decay schedule.\footnote{The linear decay was set to $10^{-4}\,$(Initial learning rate)\,/\,250\,(maximum allowed number of epochs)}. From the 165,000 synthetic observations, 70\% were used for training, 20\% for validation, and 10\% for testing. Training proceeded in epochs, with mini-batches of 128 samples per iteration. After each epoch, model performance was evaluated on the validation set, and the resulting validation score was used to track convergence and mitigate overfitting.

To improve training stability and efficiency, we employed three callback strategies. First, the learning rate was automatically reduced by a factor of 0.8 when the validation loss did not improve for five consecutive epochs. Secondly, we saved the model weights corresponding to the lowest validation loss to prevent overfitting. Thirdly, we stopped training pre-emptively when the validation loss did not improve for 20 consecutive epochs. The full training run required approximately four hours on a GPU cluster with an NVIDIA GeForce RTX 2070 Ti and about 12 hours on an Apple Silicon M2 processor.

Given the dimensionality of the output space, we evaluated the network performance as a function of the training-set size. We constructed randomly selected subsets of the full training set with sizes ranging from 20\,000 to 165\,000, and trained identical networks on each subset. We evaluated the final loss reached by the CNN, finding that it steadily decreases with increasing training size until $\sim$125\,000, beyond which increasing the size of the training sample does not significantly improve the predictions (see Fig.~\ref{fig:loss_function} in Appendix~\ref{Appendix:Loss}). This behaviour indicates that the adopted training-set size is sufficient to constrain the model and mitigates concerns about overfitting driven by an undersampled training distribution. 

Once training is complete, we evaluate the performance of the CNN model using the independent test dataset (see Sec. \ref{sec:Results}).

\subsection{\texttt{pPXF} analysis}\label{sec:pPXF}
We run {\tt pPXF} \citep{Cappellari2012, Cappellari2023} (v9.4.2) on the same test dataset used for evaluating the CNN model. {\tt pPXF} is a widely adopted method for deriving SFHs and age-metallicity relations by minimising the residuals between an observed spectrum and a linear combination of SSP templates through a non-linear least-squares optimisation. Its most recent implementation allows the simultaneous fitting of spectra and photometry \citep{Cappellari2023}, enabling a direct comparison with our network under identical input conditions and using the same SSP template set (see Sec.~\ref{sec:GenerationOfSyntheticObservations}). The fitting procedure followed the default configuration recommended for datasets of this type: a regularisation error of 1\%, a multiplicative polynomial of order 8, no additive components, and a \citet{Calzetti2000} extinction law with no constraints imposed on $A_V$.

It is important to note that {\tt pPXF} does not incorporate theoretical nebular emission predictions when determining stellar population properties. Instead, emission lines are fitted beforehand to constrain the kinematics and are removed from the spectra before the stellar population fit. Emission lines, however, are retained in the photometric fitting.

The output of {\tt pPXF} consists of V-band luminosity weights associated with the grid of SSP templates used in the fit, each characterised by a specific age and metallicity. To enable a direct comparison with the CNN model, we compute the mean metallicity at each age bin, weighted by the corresponding mass fraction.

\section{Results}\label{sec:Results}
\begin{figure*}[ht]
\centering
   \includegraphics[width=0.99\textwidth]{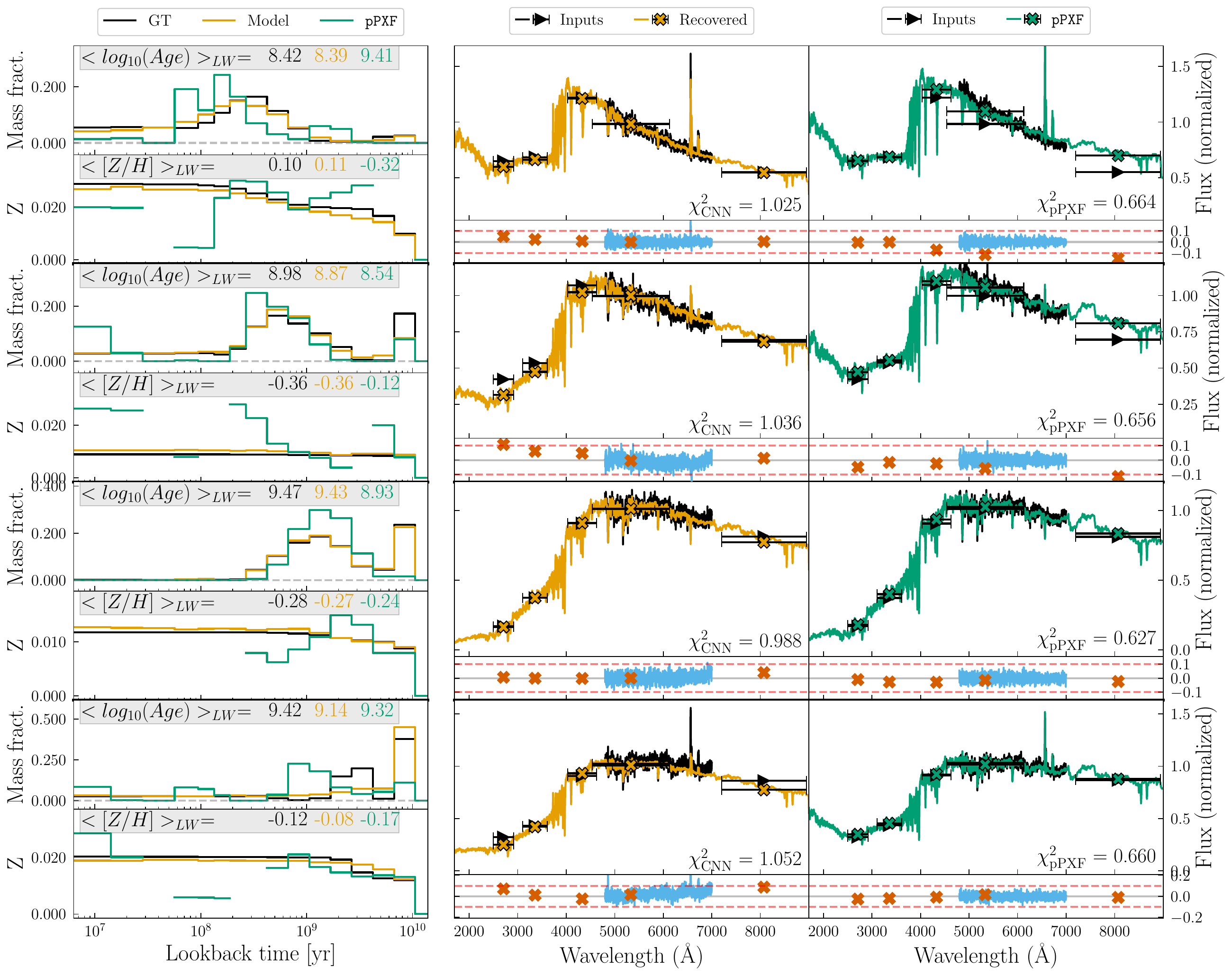} 
   \caption{Four representative examples of the recovered SFH and age--metallicity relation compared to the ground truth, using our CNN model and \texttt{pPXF}. Left column: For each example, we show the mass fraction of living stars (top panels) and the mean metallicity (bottom panels) across 16 age bins. The ground truth is shown in black, while the CNN and \texttt{pPXF} predictions are shown in orange and green, respectively. Gaps in the \texttt{pPXF} metallicity curves correspond to age bins with zero luminosity weight. 
   Central column: Normalised input spectrum (black) and CNN reconstruction (orange). Photometric fluxes in the HST/WFC3 filters F275W, F336W, F438W, F555W, and F814W are shown as symbols (ordered from shorter to longer wavelengths), with horizontal error bars indicating the effective widths of the filters. Residuals are shown in the lower panels. Right column: Same as the central column, but for the \texttt{pPXF} reconstruction (green). The values reported in the left panels correspond to the luminosity-weighted $\log(\mathrm{Age})$ and [Z/H] for the ground truth (black), CNN predictions (orange), and \texttt{pPXF} results (green). The numbers in the central and right columns indicate the $\chi^2$ values of the spectral fits.}
    \label{fig:Infered Solutions Plots}
\end{figure*}

In this section, we evaluate the predictive performance of the CNN by comparing its outputs with the ground-truth labels in the test sample and with the results obtained using the widely adopted spectral-fitting code {\tt pPXF}. Figure~\ref{fig:Infered Solutions Plots} shows four representative test cases, illustrating the SFHs and age-metallicity relations recovered by both methods. The figure also displays the input and reconstructed data generated from the solutions, enabling a direct visual assessment of the quality of the fit.

The examples shown in Fig.~\ref{fig:Infered Solutions Plots} demonstrate that the CNN is able to recover both the mass distribution of stellar populations across age bins and their corresponding metallicities with only minor deviations from the ground truth. The \texttt{pPXF} results also exhibit good overall agreement; however, in some cases, they introduce spurious contributions from very young stellar populations, often associated with metallicities that differ significantly from those of the previous components. We note that these low-level components and metallicity fluctuations are likely sensitive to the adopted regularisation scheme, extinction law, and multiplicative polynomials used in the fitting process, and they may be reduced with appropriate tuning of these parameters.

The figure also compares the input spectra with those reconstructed from the inferred solutions. Both the CNN and \texttt{pPXF} produce model spectra that closely reproduce the input data, although the $\chi^2$ of the residuals is lower for the \texttt{pPXF} solutions than for the CNN, despite the fact that the CNN more accurately recovers the ground-truth stellar population properties. We believe this behaviour arises from several methodological differences. First, \texttt{pPXF} does not include the emission lines in the fit of the stellar continuum, and it employs multiplicative polynomials to minimise differences in the continuum shape, both of which strongly affect the variance across the spectrum. More fundamentally, full spectral-fitting methods are explicitly designed to minimise the difference between the observed and fitted spectra in flux space, whereas the CNN is directly optimised to recover the underlying physical parameters. As a consequence, the network is not necessarily driven by the spectral regions with the largest flux variations, but rather by those carrying the greatest information content for the inference of stellar population properties. In the context of stellar population analysis, where strong degeneracies allow different combinations of parameters to reproduce nearly identical continuum shapes, this represents a significant advantage over traditional fitting methods.

\begin{figure}[htbp]
\includegraphics[width=0.99\linewidth]{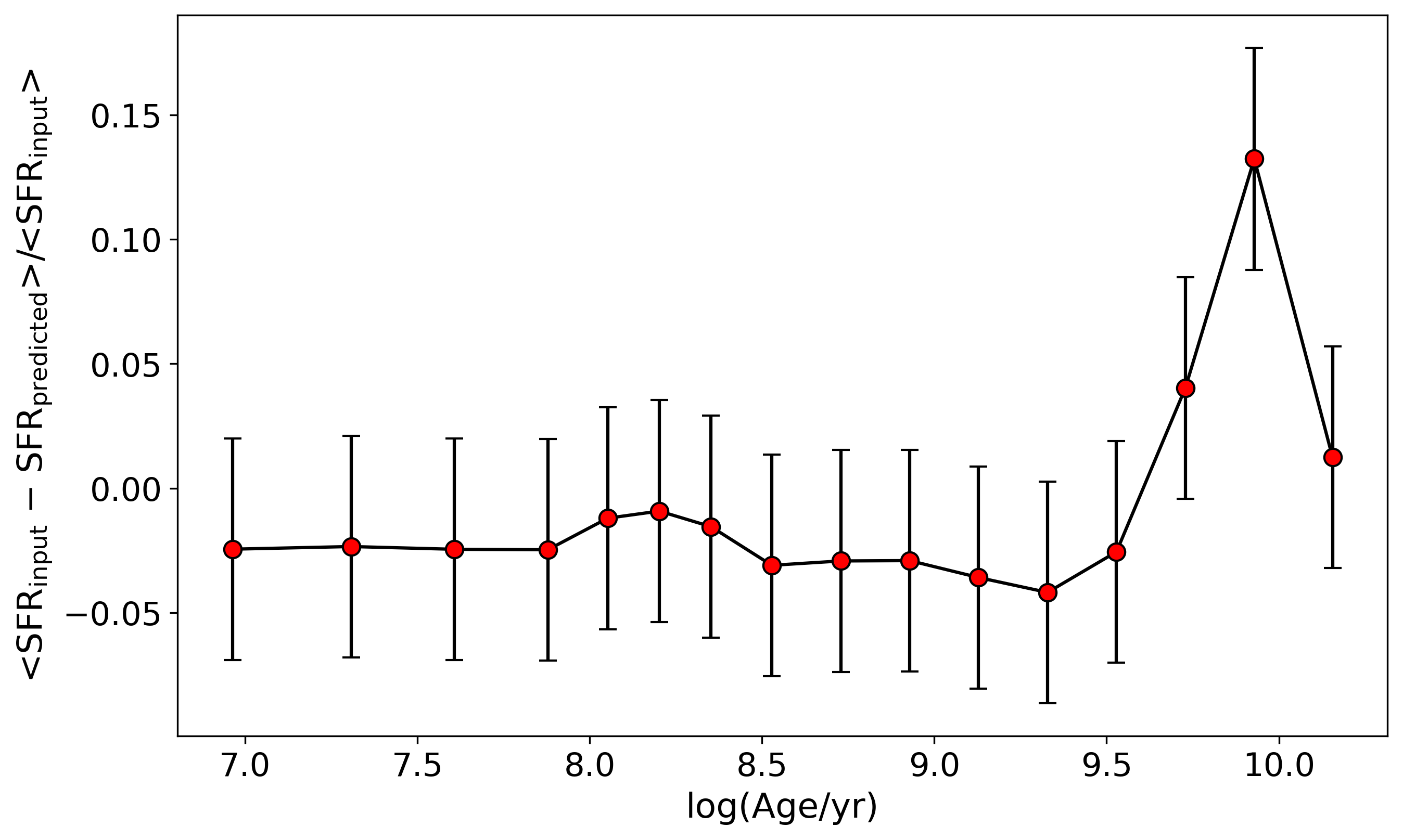}
\caption{Mean difference between the input and the predicted SFR as a function of Age, normalised by the mean SFR in each bin. The error bars represent the standard deviation against the mean.}
\label{fig:diff_SFR}
\end{figure}

Figure~\ref{fig:diff_SFR} shows the mean difference between the input and CNN-predicted SFR, normalised by the mean SFR in each age bin. The relative errors are typically on the order of 2--4\% across most age bins, increasing to $\sim$13\% in the 7--10~Gyr range. Overall, the CNN tends to overestimate the SFR at ages $\lesssim 4$~Gyr and underestimate it at older ages.

To enable a more quantitative comparison, we compute the logarithmic weighted mean ages and metallicities for both the recovered and input SFHs as follows:
\begin{equation}
\langle \log_{10}(\mathrm{age/yr}) \rangle_{\mathrm{XW}} =
\frac{\sum_i \log_{10}(\mathrm{age/yr})_i \, w_i}{\sum_i w_i},
\label{eq:weightedages1}
\end{equation}
\begin{equation}
\langle [Z/H] \rangle_{\mathrm{XW}} =
\frac{\sum_i [Z/H]_i \, w_i}{\sum_i w_i},
\label{eq:weightedages2}
\end{equation}
where $w_i$ represents either the mass fraction (for mass-weighted quantities, XW = MW) or the $V$-band light fraction (for light-weighted quantities, XW = LW) contributed by the stellar population in age bin $i$.

Although the mass-weighted age is generally a more representative indicator of the overall stellar population properties, it is less sensitive to recent star formation when most of the stellar mass resides in old populations, as is typically the case in real galaxies. By contrast, light-weighted quantities are considerably more sensitive to small contributions from young stellar populations, owing to their much lower mass-to-light ratios.

\begin{figure*}[htbp]
\centering
\includegraphics[width=0.49\linewidth,trim=4.25cm 1cm 3.5cm 2.7cm,clip]{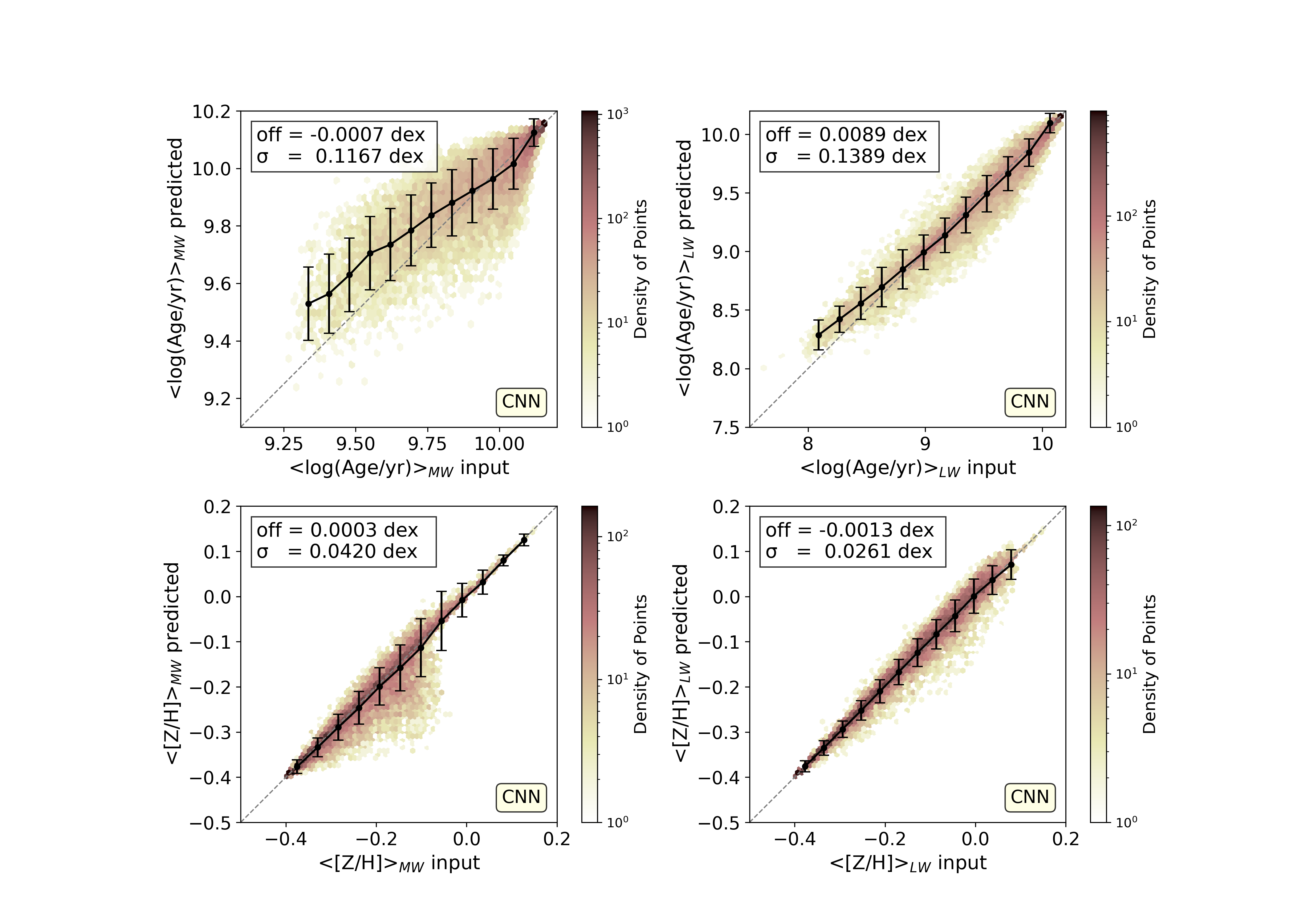}%
\includegraphics[width=0.49\linewidth,trim=4.25cm 1cm 3.5cm 2.7cm,clip]{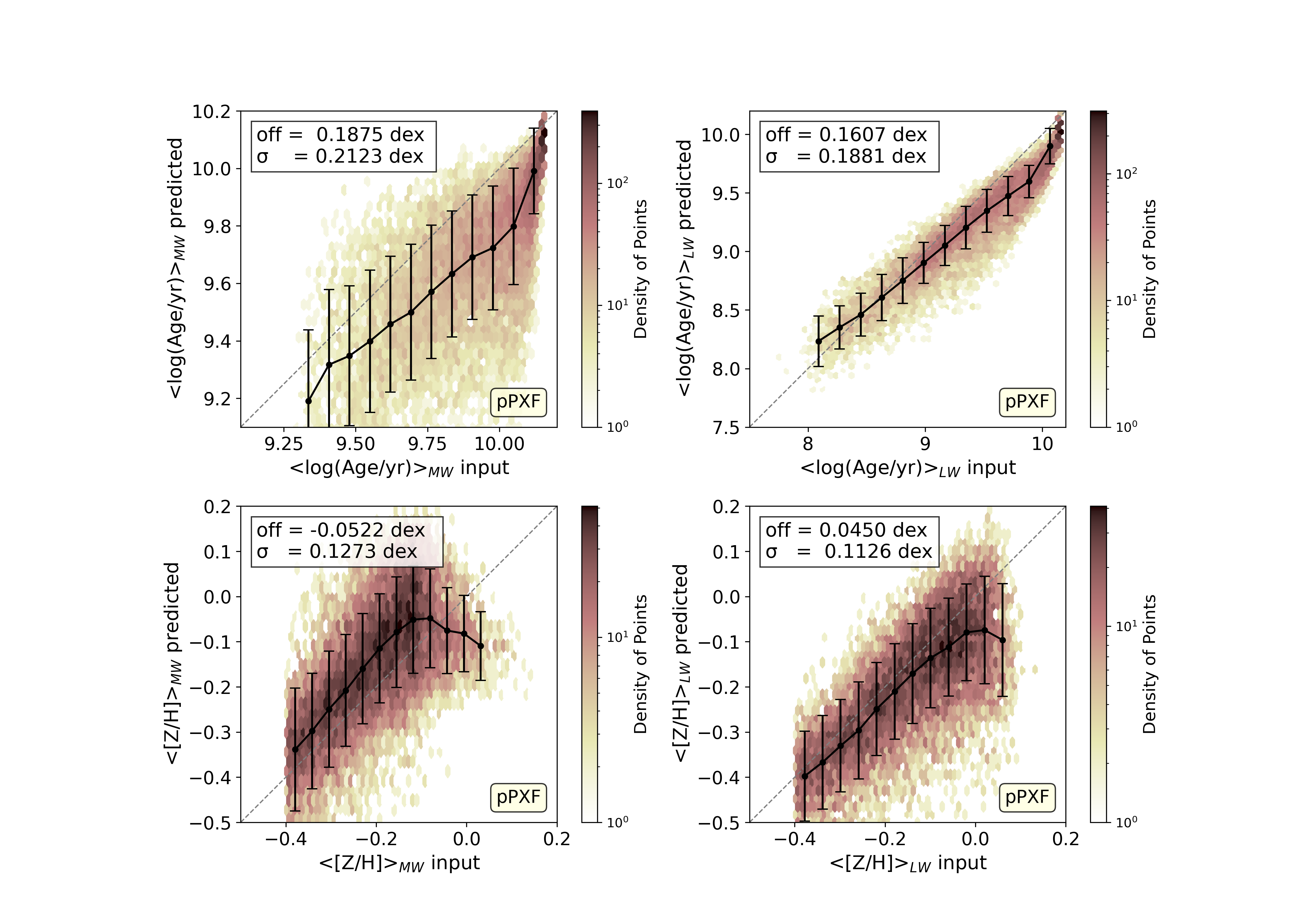}
\caption{Comparison between the ground truth mean mass- and luminosity-weighted ages (top row) and metallicities (bottom row) obtained with the CNN model (left panels) and with \texttt{pPXF} (right panels). The insets display the mean logarithmic differences between the input and predicted values ($\langle \log(\mathrm{Age/yr})_{\rm input} - \log(\mathrm{Age/yr})_{\rm predicted} \rangle$ and $\langle [Z/H]_{\rm input} - [Z/H]_{\rm output} \rangle$), together with their standard deviations. Each hexbin is coloured according to its point density, as indicated by the colour bar. Black symbols denote the median predicted values within bins of the corresponding input quantities.}
    \label{fig:comparison_agemw_hexbin}
\end{figure*}

Figure~\ref{fig:comparison_agemw_hexbin} compares the input mean ages and metallicities with those predicted by our CNN and by \texttt{pPXF}. Our CNN accurately recovers the mean values, with negligible biases in metallicity and in the mass-weighted ages, while the median bias in the light-weighted ages remains below $2\%$. The scatter in the age predictions is $30\%$ and $37\%$ for the mass- and light-weighted values, respectively, and $10\%$ and $6\%$ for the corresponding mass- and light-weighted metallicities.

In contrast, the ages recovered by \texttt{pPXF} exhibit significantly larger biases, particularly for the mass-weighted quantities, which are underestimated by approximately $65\%$ on average. The metallicity biases obtained with \texttt{pPXF} are also larger, reaching $\sim10$--$14\%$ for the mass- and light-weighted values, respectively. The scatter in the predicted quantities is likewise substantially smaller for the CNN than for \texttt{pPXF}, especially in the case of the metallicity estimates.

\subsection{Age-Z degeneracy}
\begin{figure}[t]
 \centering
\includegraphics[width=\linewidth,trim=1.5cm 1.5cm 3cm 3cm, clip]{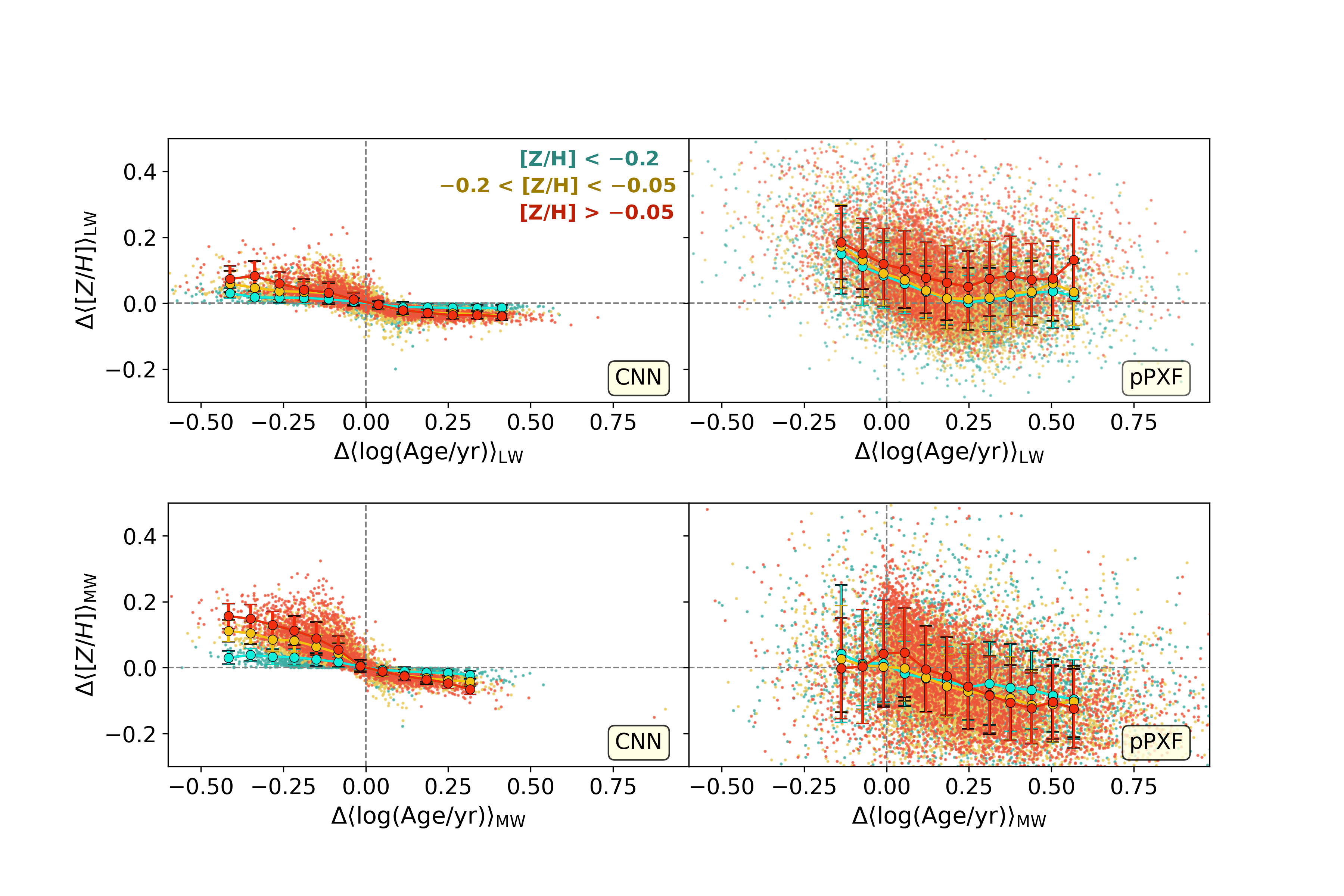}
\caption{Differences between the input and predicted mean log(age) as a function of the corresponding differences in [Z/H]. Points are colour-coded by the ground-truth luminosity-weighted metallicity as indicated in the insets. The top (bottom) panels show luminosity-weighted (mass-weighted) quantities. Left panels display the results obtained with the CNN, while the right panels show those derived using  {\tt pPXF}.}
 \label{fig:deg_age_Z}
\end{figure}

To investigate the impact of the age-metallicity degeneracy on the predictions of both methods, we show in Fig.~\ref{fig:deg_age_Z} the correlation between the differences in the logarithm of the input and recovered mean ages for both mass- and luminosity-weighted estimates, as predicted by the CNN model and \texttt{pPXF}.
As can be seen, there is an anti-correlation in the errors of both parameters, which are more readily seen in the mass-weighted values.

However, the model shows a significantly shallower slope of the correlation, much less scatter overall, and the distribution is better centred around $\Delta\!=\!0$ in both axes. {\tt pPXF} shows a systematic shift towards underestimating ages regardless of how the age is weighed. The metallicities, on the other hand, are overestimated when weighting by mass and the opposite when weighting by light (as seen also in Fig. \ref{fig:comparison_agemw_hexbin}). This indicates that {\tt pPXF} systematically warps the shape of the metallicity evolution and overestimates recent SFR, on top of providing less accurate values in general. Another interesting difference between the model and {\tt pPXF} is how the value of the metallicity correlates with the scatter; no significant gradient is observed for {\tt pPXF}, but it is clearly present for the model, such that lower values of the metallicity are better determined, especially when the age is overestimated. Older stellar populations are expected to have deeper metal absorption features, so it makes sense that very low metallicities are better predicted when the model overestimates the age, since in this case, the absorption features are faint in the first place.

These results suggest that our CNN is able to capture the subtle differences in the spectra and flux values resulting from combinations of ages and metallicities that are difficult to capture using full spectral fitting methods such as \texttt{pPXF}. This advantage of CNNs has also been demonstrated in \cite{Woo2024}, where they compare several fitting codes with the deep-learning CNN \texttt{STARNET} \citep{Fabbro2018}.

\begin{figure}[ht]
\includegraphics[width=\linewidth]{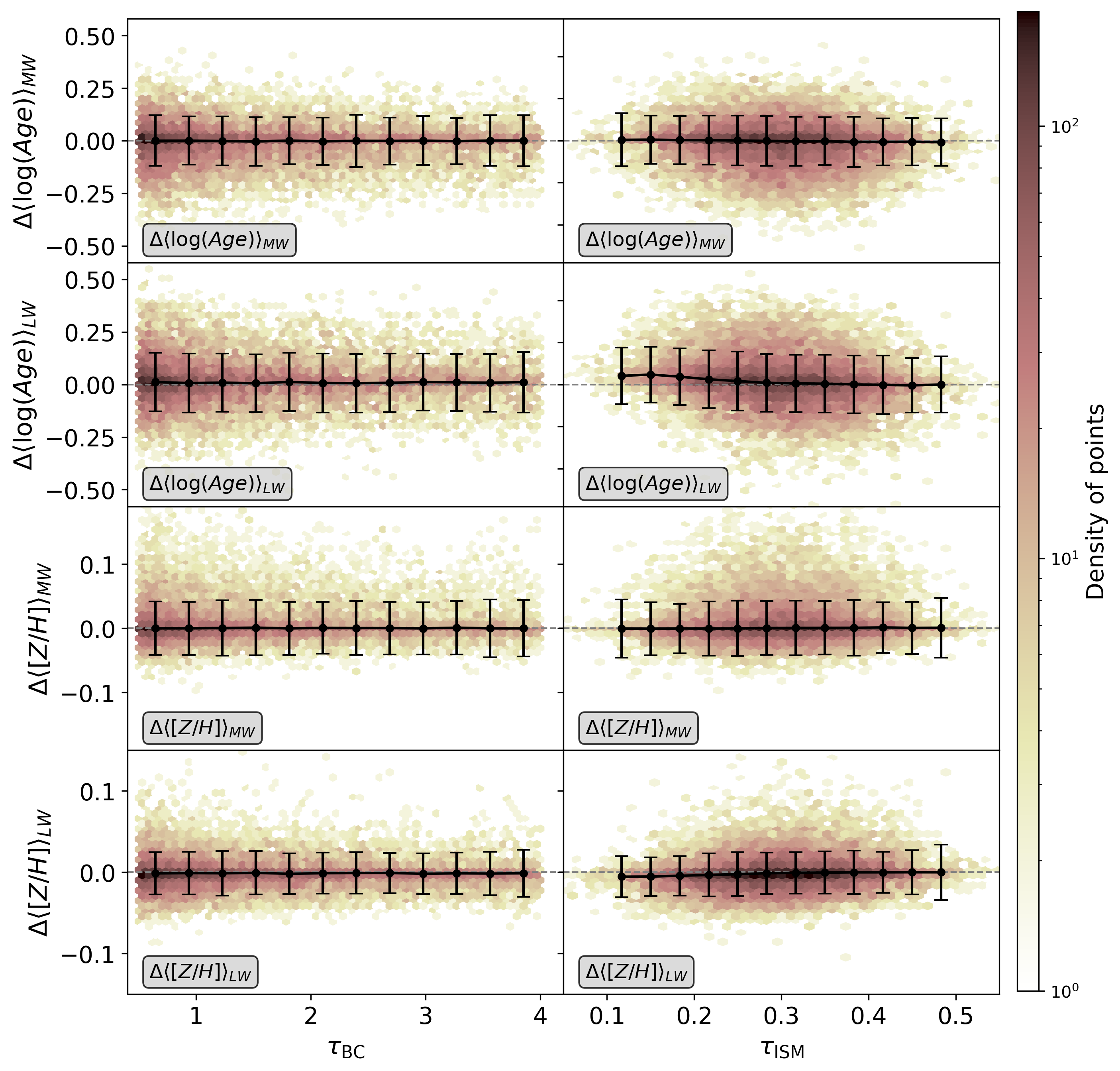}
\caption{Density plots of the differences between the ground-truth and predicted mean luminosity- and mass-weighted ages and metallicities as a function of the optical depth of the birth cloud ($\tau_{\rm BC}$; left panels) and the diffuse interstellar medium ($\tau_{\rm ISM}$; right panels). The black symbols indicate the median differences computed in bins of $\tau$, while the error bars represent the corresponding standard deviations.}
\label{fig:diff_taus}
\end{figure}

 Although the synthetic spectra are attenuated using a transfer function (see Sec.~\ref{sec:GenerationOfSyntheticObservations}) to mimic dust extinction, the CNN is not trained to predict the corresponding optical depths. It is therefore important to evaluate whether the amount of extinction affects the recovered parameters. Figure~\ref{fig:diff_taus} presents the differences between the ground-truth and predicted mean luminosity- and mass-weighted ages and metallicities as a function of the optical depth of the birth cloud ($\tau_{\rm BC}$) and the diffuse interstellar medium ($\tau_{\rm ISM}$). As shown in the figure, the biases in the predicted mean parameters do not exhibit any significant correlation with the level of extinction in either dust component included in the synthetic data. We also check that the network preserves the age-metallicity correlations present in the input data,  indicating that the network does not wash out the underlying correlations, despite the introduction of a broad range of extinction and noise (see Fig.\ref{fig:Age_Z_correlations} on Appendix\ref{sec:Age_Z_correlations}).

\subsection{Effect of the SNR in the predictions}\label{sec:Validation}

\begin{figure}[!htbp]
    \centering
    \includegraphics[width=0.99\linewidth]{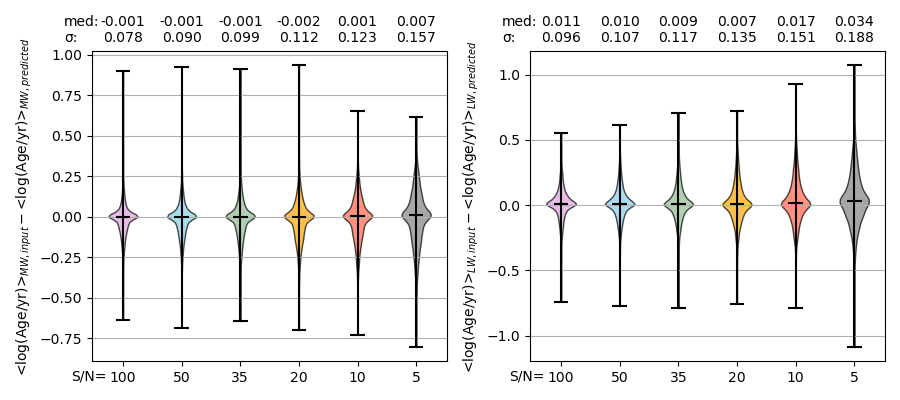}\\
    \includegraphics[width=0.99\linewidth]{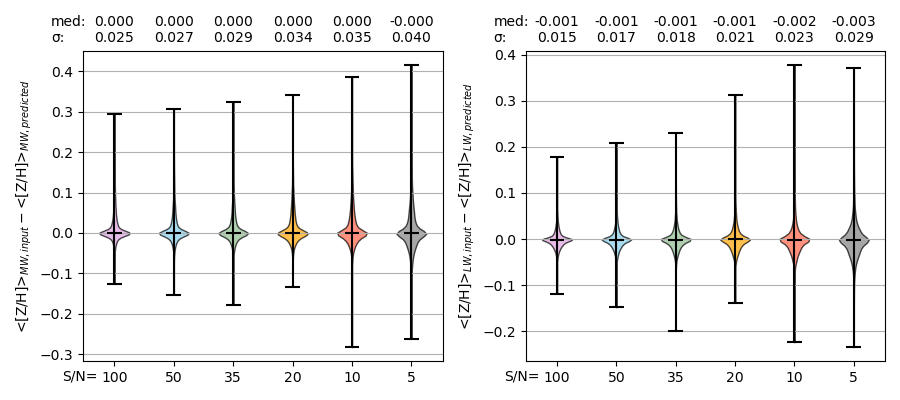}
    \caption{Violin plots showing the distributions of the differences between the input and predicted mean values of the MW and LW age and metallicities for different SNR levels. The black horizontal line marks the median, and the error bars indicate the 16th and 84th percentiles of each distribution. The median values and standard deviations are annotated at the top of each panel.}
    \label{fig:ruido_snr}
\end{figure}

To assess the robustness of the predictions depending on the level of spectral noise, we evaluated the test sample after adding Poisson noise corresponding to six different SNRs (SNR = 5--100) with the model. Figure~\ref{fig:ruido_snr} presents the median differences and associated scatter between the input and predicted values as a function of SNR.

As expected, the scatter in the recovered parameters increases toward lower SNR, approximately doubling between SNR = 100 and SNR = 5. In contrast, the impact on the mean bias is more modest: although the bias increases by a few per cent at low SNR, it remains small overall. The clear dependence of the scatter on SNR provides an important indication that the CNN is not overfitting the training data, but instead is extracting information from the underlying signal.

Real spectra may include additional sources of systematic noise arising, for example, from residuals of imperfect sky-line subtraction or bad pixels. To investigate the impact of such effects on our results, we also generated synthetic data using error spectra derived from the observations. These were constructed from the root mean squared of the residuals of a fit to the stellar and gas kinematics obtained with {\tt pPXF}. We run the model trained with Poisson noise on the test set, where we added noise using this ``realistic'' error spectra, altering each pixel individually according to this error spectra. The results, which have been included in Appendix~\ref{app:realistic_noise}, are consistent with those obtained on the test sample with Poisson noise. These tests demonstrate that our noise-injection prescription does not bias the model's predictions, likely because the sky residuals do not correlate with any of the desired outputs, and the network ignores the inputs.

\subsection{Ablation analysis of the input data}\label{sec:Input dependencies}

Identifying which input features carry the greatest predictive power provides insight into the spectral regions and photometric bands that most strongly constrain the SFH and the age-metallicity relation. To investigate this, we perform an ablation analysis in which the fluxes in selected spectral regions are replaced, pixel by pixel, with values randomly drawn from the empirical distribution of the same region across the full dataset. This procedure follows the permutation feature-importance approach introduced by \citet{Breiman2001} for random-forest models and later formalised and generalised by \citet{Fisher2018_published}, and has been successfully applied in related contexts \citep[e.g.][]{Kang2023}. By construction, this method removes any association between the perturbed feature and the target variables while preserving the global statistical properties of the data.

We quantify the effect of removing individual HST fluxes on the bias and scatter of the predicted mean ages and metallicities, and we also consider the limiting case in which no photometric information is provided.

We further assess the sensitivity of the CNN predictions to different spectral regions by replacing selected wavelength intervals in the input spectra. We test contiguous equal-length intervals of width $\Delta\lambda \simeq 250$~\AA{} covering the full wavelength range, as well as the regions containing H$\alpha$ (6500--6750~\AA{}) and H$\beta$ (4800--5050~\AA{}). To evaluate the importance of the continuum shape, we also consider ``flattened'' spectra, obtained by fitting and subtracting a second-order polynomial.

\subsubsection{Ablation tests on the age predictions}
\label{sec:photometricfluxes}

\begin{figure*}
    \centering
    \includegraphics[width=0.99\linewidth]{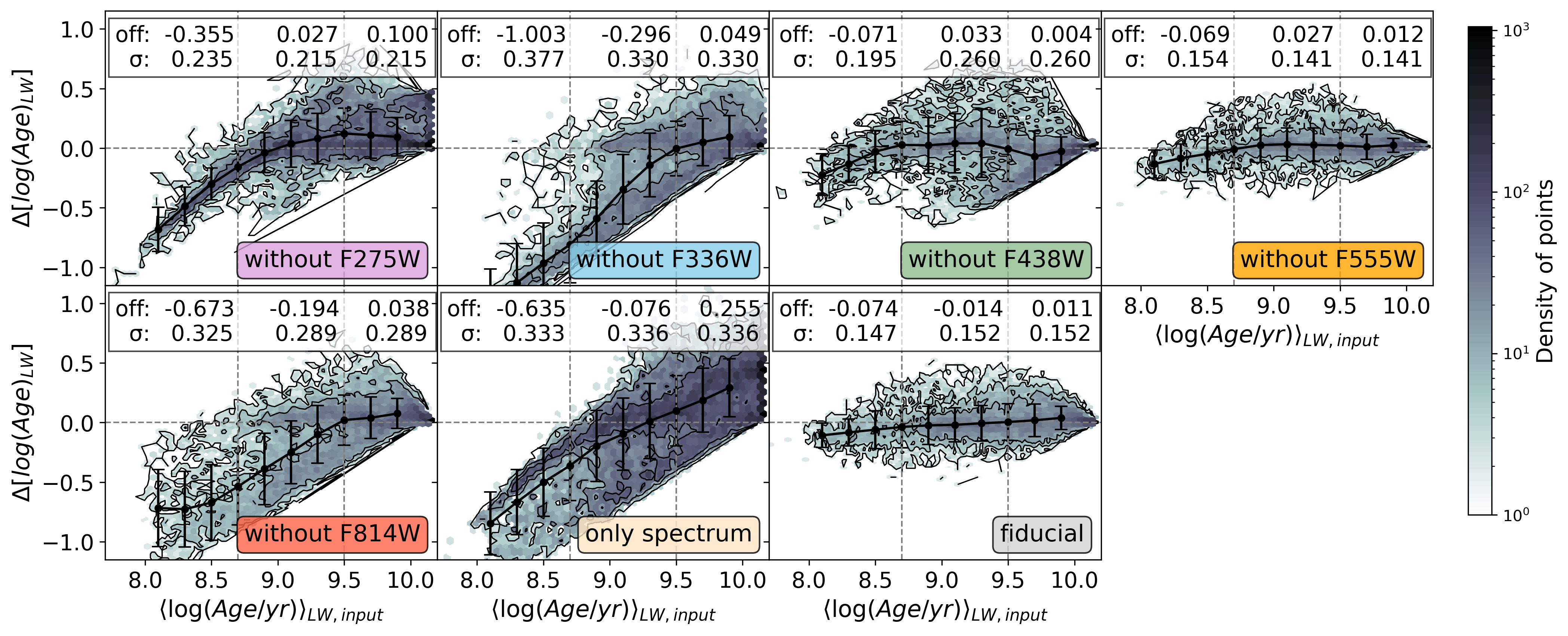}
    \caption{Impact of removing photometric fluxes from the input data on the recovery of the light-weighted age. Each panel shows the difference between the predicted and input light-weighted age as a function of the input value, when individual photometric fluxes are replaced by values randomly drawn from the empirical distribution of the whole sample. Additionally, the full photometric input can be replaced with random values, leaving only the spectrum unchanged. For reference, the fiducial case in which all photometric fluxes are included is also shown. The running median and standard deviation are plotted as black circles with error bars. The insets report the mean bias and standard deviation in three age intervals, whose boundaries are marked by vertical lines: (1) $\langle \log(\mathrm{Age}/\mathrm{yr}) \rangle_{\rm LW} \leq 8.6$; (2) $8.6 < \langle \log(\mathrm{Age}/\mathrm{yr}) \rangle_{\rm LW} \leq 9.5$; and (3) $\langle \log(\mathrm{Age}/\mathrm{yr}) \rangle_{\rm LW} > 9.5$.}
    \label{fig:ablation_age_lw_phot}
\end{figure*}

Figure~\ref{fig:ablation_age_lw_phot} quantifies the impact of removing individual photometric bands on the recovery of the mean light-weighted (LW) age as a function of the input mean age. The fiducial case, in which all photometric fluxes are included, is shown for comparison. Overall, removing individual HST bands degrades the predictions, introducing systematic offsets that strongly depend on the mean age of the system.

For populations with $\langle \log(\mathrm{Age}/\mathrm{yr}) \rangle_{\rm LW} \leq 8.6$, the largest degradation is produced by removing F336W, followed by F275W and F814W. In the first case, the bias increases from 15\% in the fiducial case to 100\%, while removing F275W and F814W increases it to 80\% and 50\%, respectively. By contrast, removing F438W or F555W does not produce a significant increase in the bias for these young systems.

The impact on intermediate-age populations ($8.6 < \langle \log(\mathrm{Age}/\mathrm{yr}) \rangle_{\rm LW} \leq 9.5$) is smaller, but still significant. In this regime, the bias in the mean LW age increases from $3\%$ in the fiducial case to $50\%$ and $35\%$ when the F336W and F814W bands are removed, respectively. The age predictions of older systems ($\langle \log(\mathrm{Age}/\mathrm{yr}) \rangle_{\rm LW} > 9.5$) depend less strongly on the photometric fluxes, with the notable exception of F275W. The omission of this band produces an average bias of 0.2~dex, together with an increase in the scatter of more than $100\%$ relative to the fiducial case.

Removing all photometric fluxes also degrades the predictions substantially. In this case, the network systematically under-predicts the ages of young systems and over-predicts those of old ones, with biases reaching up to 75\%. This demonstrates that the photometric information provides an essential anchor for the age scale across the full parameter space.

\begin{figure*}
    \centering
    \includegraphics[width=0.99\linewidth]{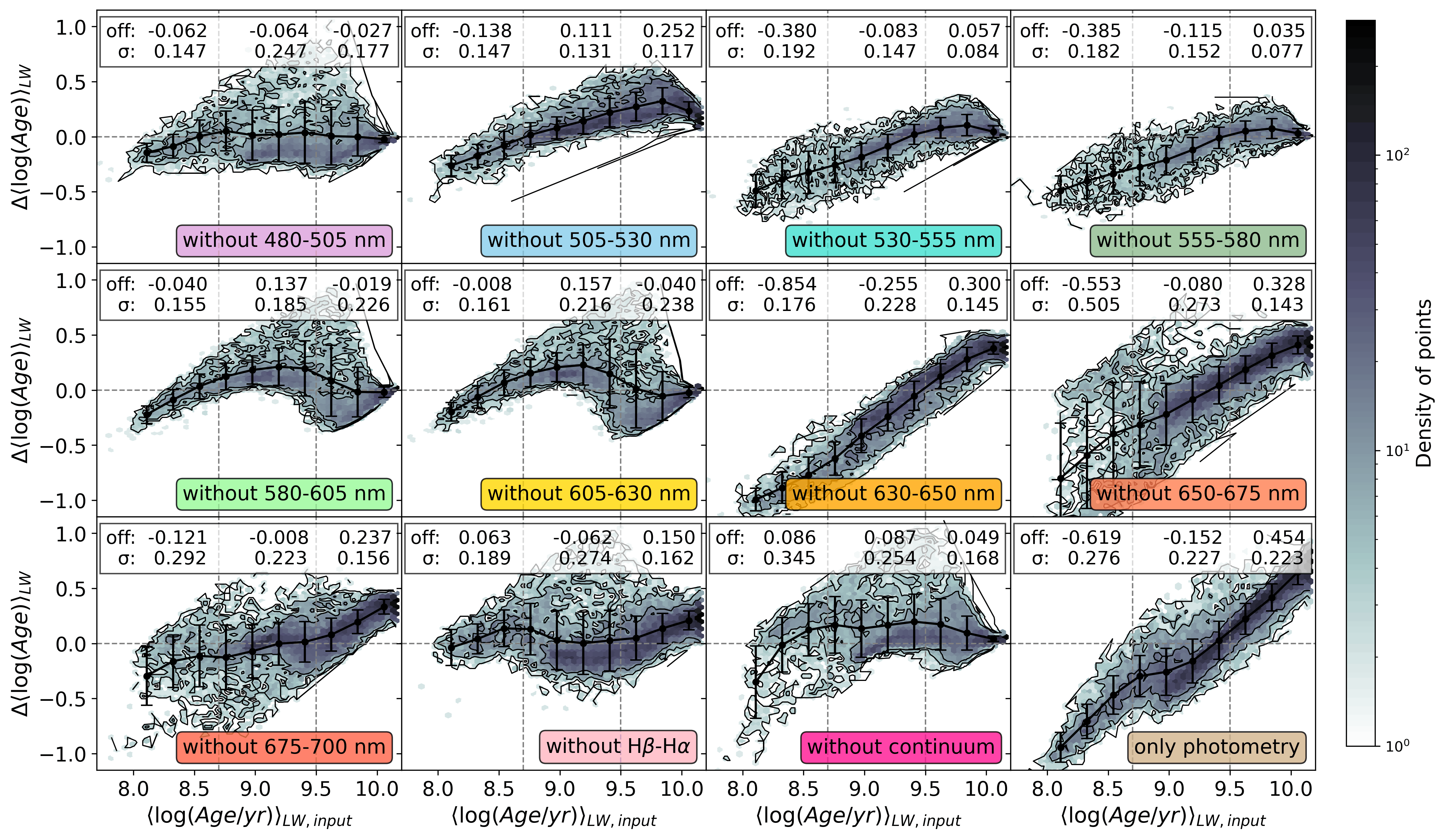}
    \caption{Impact of different spectral regions on the LW-age predictions. Each panel shows the difference between the ground-truth and predicted $\langle \log(\mathrm{Age}/\mathrm{yr}) \rangle_{\rm LW}$ for different perturbations of the input spectrum. The first nine panels (from left to right and top to bottom) show the results obtained when individual wavelength segments are replaced by random values drawn from the empirical distribution of the same segment across the full dataset. The panel labelled ``without continuum'' shows the results when the continuum shape is removed from the spectra. The final panel shows the case in which the full spectrum is replaced by random values. The running median and standard deviation are plotted as black circles with error bars. The insets report the mean bias and standard deviation in three age intervals, whose boundaries are marked by vertical lines: (1) $\langle \log(\mathrm{Age}/\mathrm{yr}) \rangle_{\rm LW} \leq 8.6$; (2) $8.6 < \langle \log(\mathrm{Age}/\mathrm{yr}) \rangle_{\rm LW} \leq 9.5$; and (3) $\langle \log(\mathrm{Age}/\mathrm{yr}) \rangle_{\rm LW} > 9.5$.}
    \label{fig:violin_means_spectra_secciones}
\end{figure*}

Figure~\ref{fig:violin_means_spectra_secciones} illustrates the impact of perturbing different spectral regions on the mean LW-age predictions. Overall, the inferred values show a strong dependence on the integrity of the input spectrum, with the replacement of individual wavelength intervals systematically increasing both the residual scatter and the bias. As in the case of the photometric ablation tests, the magnitude of the bias also depends strongly on the age of the system.

For the youngest systems, the strongest degradation occurs when removing the 6300--6500~\AA{} and 6500--6750~\AA{} intervals, where the median residuals can shift by up to $\sim0.5$~dex. These wavelength ranges contain several strong emission lines, including H$\alpha$, [N\,{\sc ii}], and [S\,{\sc ii}], which are associated with young ionising stellar populations. However, removing regions without strong emission lines, such as 5300--5550~\AA{} and 5550--5800~\AA{}, also degrades the age predictions for young systems by up to 0.4~dex. These intervals contain weak metal-blend absorption features, including the Lick indices Fe5335, Fe5406, Fe5709, and Fe5782, which become significantly stronger in populations aged $0.2$--$1$~Gyr \citep[e.g.][]{Vazdekis2010}.  Furthermore, the continuum slope in the 5300--5550~\AA{} segment, which corresponds to the normalisation range adopted for the input spectra, shows very little variation even among spectra with substantially different stellar populations. This indicates that the network does not rely exclusively on large-scale spectral variations or strong emission lines, but also exploits weaker absorption features in the determination of young stellar ages.

Intermediate-age and old populations show a similar trend: the strongest degradation again occurs when the 6300--6500~\AA{} and 6500--6750~\AA{} regions are removed, with the offset increasing by up to $90\%$ relative to the fiducial case. In addition to H$\alpha$ absorption, these wavelength ranges include molecular bands and Fe lines whose strengths depend on both age and metallicity. Similarly, removing the 5050--5300~\AA{} interval also introduces strong biases of up to $\sim80\%$ in the predicted LW ages. This spectral range contains some of the strongest metallicity-sensitive absorption features, such as Mgb and Fe5270, and, together with the 6300--6500~\AA{} segment, has one of the largest impacts on the metallicity predictions (see Sect.~\ref{sec:spectralregions}). The degradation caused by removing these regions likely reflects a poorer recovery of the metallicity, which in turn propagates into the age estimates for old systems, where the age-metallicity degeneracy is particularly strong.

Although the photometric fluxes (only photometry) have a strong influence on the predicted values, removing all the spectral information produces the largest degradation in the age predictions across all age ranges. The photometric fluxes are particularly important for young systems since the MUSE spectral range does not extend sufficiently far into the blue, where sensitivity to young stellar populations is greatest. The spectrum, however, remains essential for accurate age determination across the full range of stellar populations. In particular, spectral regions containing strong absorption features, such as Mgb, TiO molecular bands, and prominent Fe lines, play a major role in determining the ages of old systems.

Furthermore, flattening the input spectra (without continuum) results in a systematic over-prediction of the mean LW age, although the offset is not strongly correlated with the mean age itself. This suggests that the model is not only sensitive to age-dependent variations in the continuum shape of the spectra, but that detailed absorption features also remain essential for obtaining physically meaningful results.

The results of the ablation tests for the mass-weighted (MW) ages are shown in Figs.~\ref{fig:ablation-age-mw-phot} and~\ref{fig:ablation-age-mw-spec} of Appendix~\ref{app:ablation_metallicity}. Overall, the conclusions are similar to those for the LW predictions, although the degradation is smaller. The most influential bands for the age predictions of young and intermediate-age populations are again F336W and F814W. However, the bias resulting from removing F275W is much lower for MW ages, as expected, since the NUV band primarily constrains very young populations, which usually contribute little to the total stellar mass.
More generally, the MW ages depend less strongly on the global shape of the SED than the LW values, which suggests that the MW ages rely more on spectral features and less on the overall shape of the spectrum than the LW-ages.


\subsubsection{Ablation tests on metallicity predictions}
\label{sec:spectralregions}

\begin{figure*}[ht]
    \centering
     \includegraphics[width=17cm]{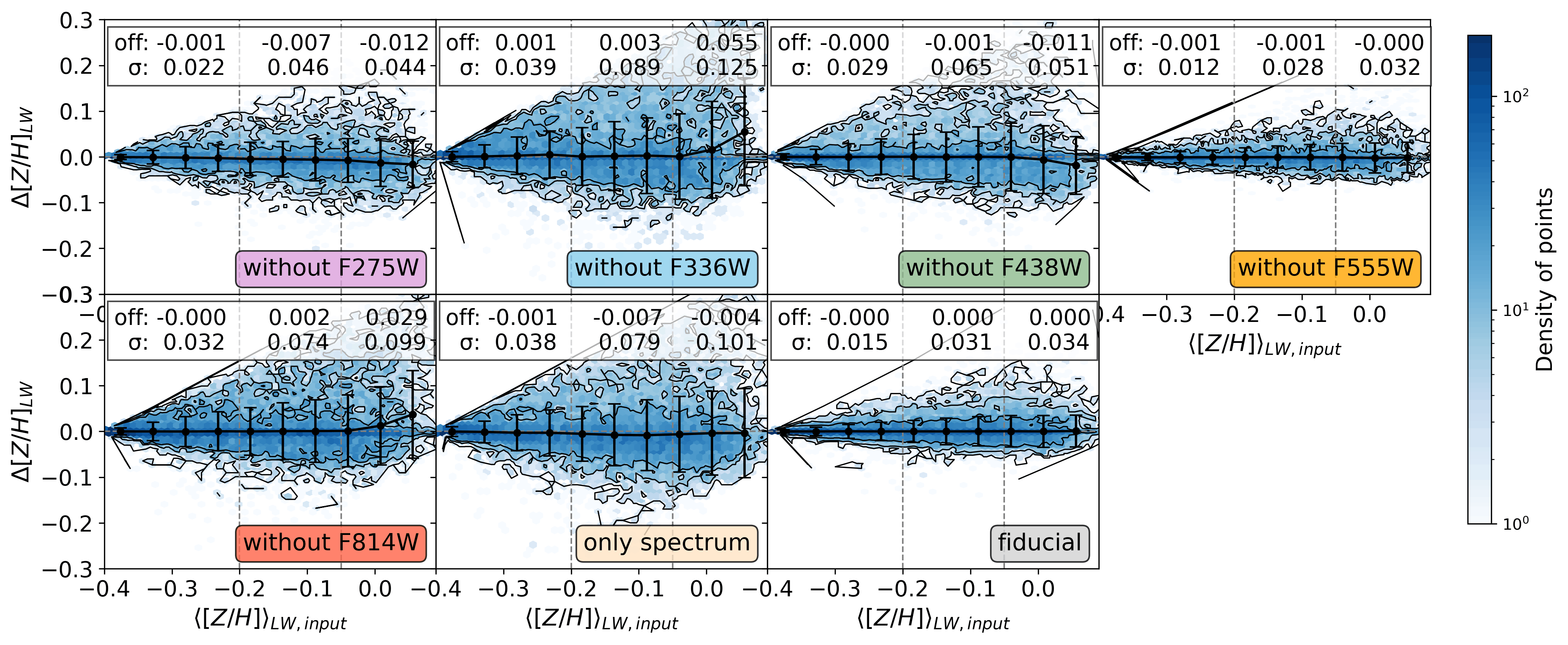}
 \caption{Same as Fig.~\ref{fig:ablation_age_lw_phot}, but for the light-weighted metallicity. The insets report the mean bias and standard deviation in three metallicity intervals, whose boundaries are marked by vertical lines: (1) $\langle [Z/H] \rangle_{\rm LW} \leq -0.2$; (2) $-0.2 < \langle [Z/H] \rangle_{\rm LW} \leq 0.05$; and (3) $\langle [Z/H] \rangle_{\rm LW} > 0.05$.}
    \label{fig:ablation-z-phot0}
\end{figure*}

Figure~\ref{fig:ablation-z-phot0} shows the impact of the photometric fluxes on the mean LW metallicity predictions. Unlike in the age case, the accuracy of the predictions does not show a clear dependence on the input metallicity and appears largely insensitive to the removal of individual filters.

However, removing individual bands does increase the scatter, particularly for metal-rich systems. The filters with the largest impact are, in decreasing order, F336W, F814W, F438W, and F275W, whereas removing F555W does not produce any significant increase in either the bias or the scatter of the predictions. The effect is strongest for systems with intermediate or high metallicity ($\langle [Z/H] \rangle > -0.2$), where removing F336W and F814W increases the scatter in the mean LW metallicity by $15\%$ and $12\%$, respectively, relative to the fiducial case. In contrast, the scatter in low-metallicity systems is much less affected, with a maximum increase of only $4\%$. Since the synthetic data include metallicity evolution, metal-poor systems are typically older than metal-rich ones. We therefore interpret this behaviour as a consequence of the lower sensitivity of the photometric fluxes, relative to the spectral features, in constraining the properties of these older stellar populations (see Sect.~\ref{sec:photometricfluxes}).

\begin{figure*}[ht]
    \centering
     \includegraphics[width=17cm]{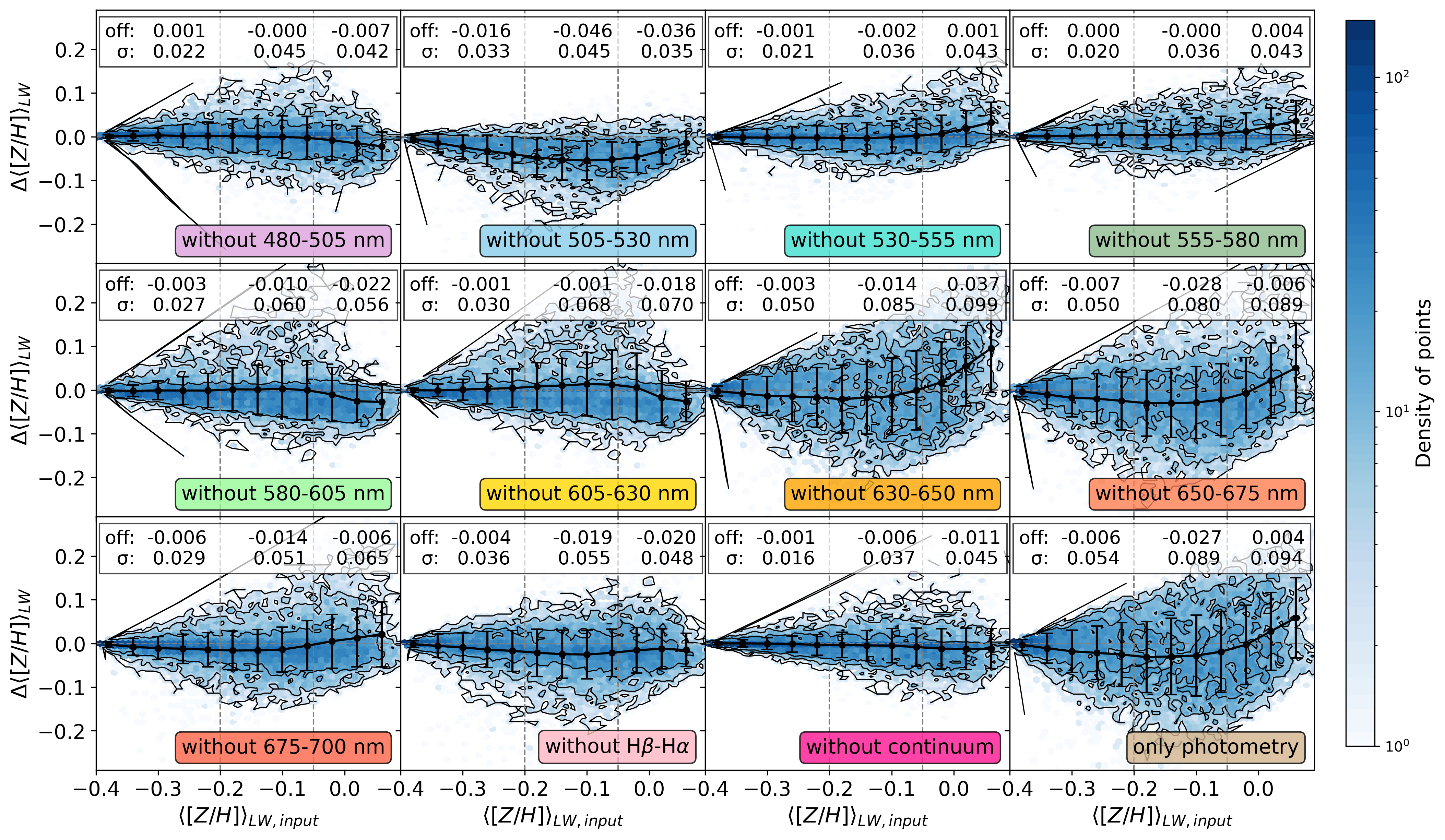}
 \caption{Impact of different spectral regions on the predictions of the mean LW metallicity. The different panels are equivalent to those in Fig.~\ref{fig:violin_means_spectra_secciones}, but for $\langle [Z/H] \rangle_{\rm LW}$. Masking most wavelength intervals produces only minor changes in accuracy, while removing broad continuum information or relying on photometry alone leads to larger scatter, demonstrating the complementary role of spectral and photometric data in constraining metallicity. The insets report the mean bias and standard deviation in three metallicity intervals (see Fig.~\ref{fig:ablation-z-phot0}).}
\label{fig:ablation-z-spec0}
\end{figure*}

Figure~\ref{fig:ablation-z-spec0} shows the impact of the spectral ablation tests on the mean LW metallicity predictions. As in the photometric tests, the metallicity predictions do not show strong systematic biases when individual spectral regions are removed, regardless of the mean metallicity of the system. However, the scatter increases clearly with increasing metallicity. The largest degradation occurs when the 6300--6500~\AA{} and 6500--6750~\AA{} intervals are removed, producing a $12\%$ increase in the scatter relative to the fiducial case for systems with $\langle [Z/H] \rangle_{\rm LW} > -0.2$. This behaviour reflects the importance of strong emission lines, such as [N\,{\sc ii}], [S\,{\sc ii}], and H$\alpha$, in constraining the metallicity of young, metal-rich systems. The next most influential intervals are 6050--6300~\AA{} and 5800--6050~\AA{}, whose removal increases the scatter by about $9\%$ in the same metallicity regime. The remaining spectral segments produce smaller increases, typically between $3\%$ and $5\%$.

The equivalent figures for MW metallicities are presented in Figs.~\ref{fig:ablation-z-phot} and~\ref{fig:ablation-z-spec} of Appendix~\ref{app:ablation_metallicity}. The results are very similar to those for the LW values, with only small numerical differences. The main discrepancy is the reduced scatter in the high-metallicity regime, likely due to the smaller number of synthetic spectra with high mean metallicity.

\section{Application to PHANGS-MUSE data: NGC 3627}\label{sec:NGC3627}

In this section, we test the performance of our trained CNN using real observational data of the nearby spiral galaxy NGC\,3627. The aim is to assess the model's performance on real data and to investigate the presence of possible domain biases.

We compare the results from our network with those obtained using the PHANGS-MUSE Data Analysis Pipeline (\texttt{DAP}\footnote{The version of the pipeline used to analyse the PHANGS-MUSE galaxies is publicly available at \url{https://gitlab.com/francbelf/ifu-pipeline}.}; Em22), which performs full spectral fitting with \texttt{pPXF} on the MUSE spectroscopic data without incorporating HST photometry. The spectra are Voronoi-binned to ensure a minimum signal-to-noise ratio of $\mathrm{SNR} \sim 35$ per\,\AA{}.

The public data release includes mass- and luminosity-weighted ages and metallicities derived using the E-MILES simple stellar population models \citep{Vazdekis2016}, generated with the BaSTI isochrones \citep{Pietrinferni2004}, with an age coverage from 0.15 to 14~Gyr. However, for a more consistent comparison with our results, we use values from an internal version of the PHANGS-MUSE analysis that adopts the same set of stellar population templates used in this work, based on E-MILES extended to very young ages (6.3~Myr). Nevertheless, the PHANGS-MUSE implementation does not include templates older than 14.12~Gyr or with metallicities below [Z/H] $< -0.7$, which introduces differences in the accessible parameter space relative to our model.

We use the same Voronoi-binned spectra as input for our CNN, preprocessed in the same way as the synthetic spectra used for training. In particular, all spectra are shifted to the rest frame, broadened to a final velocity dispersion of $\sigma = 150$~km~s$^{-1}$, and normalised by their mean flux in the wavelength range 5300--5500\,\AA. For the first two steps, we use the radial velocities and velocity dispersions derived for each bin by the PHANGS-MUSE \texttt{DAP}.

To use the HST fluxes as input to our CNN, we first match the imaging data to the same spatial binning scheme as the spectra. This is achieved by first converting the images to surface brightness, convolving them to the MUSE point-spread function, and then averaging the fluxes within each Voronoi bin to obtain the corresponding photometric measurements. We only retain bins for which all spatially averaged fluxes are positive.

The preprocessed spectra and photometry are then fed into the trained CNN to infer the SFHs and age-metallicity relations in the predefined age bins. From these outputs, we compute the mean stellar ages and metallicities, weighted by both stellar mass and luminosity, following Eqs.~\ref{eq:weightedages1} and \ref{eq:weightedages2}.

\begin{figure*}
  \resizebox{\hsize}{!}{\includegraphics{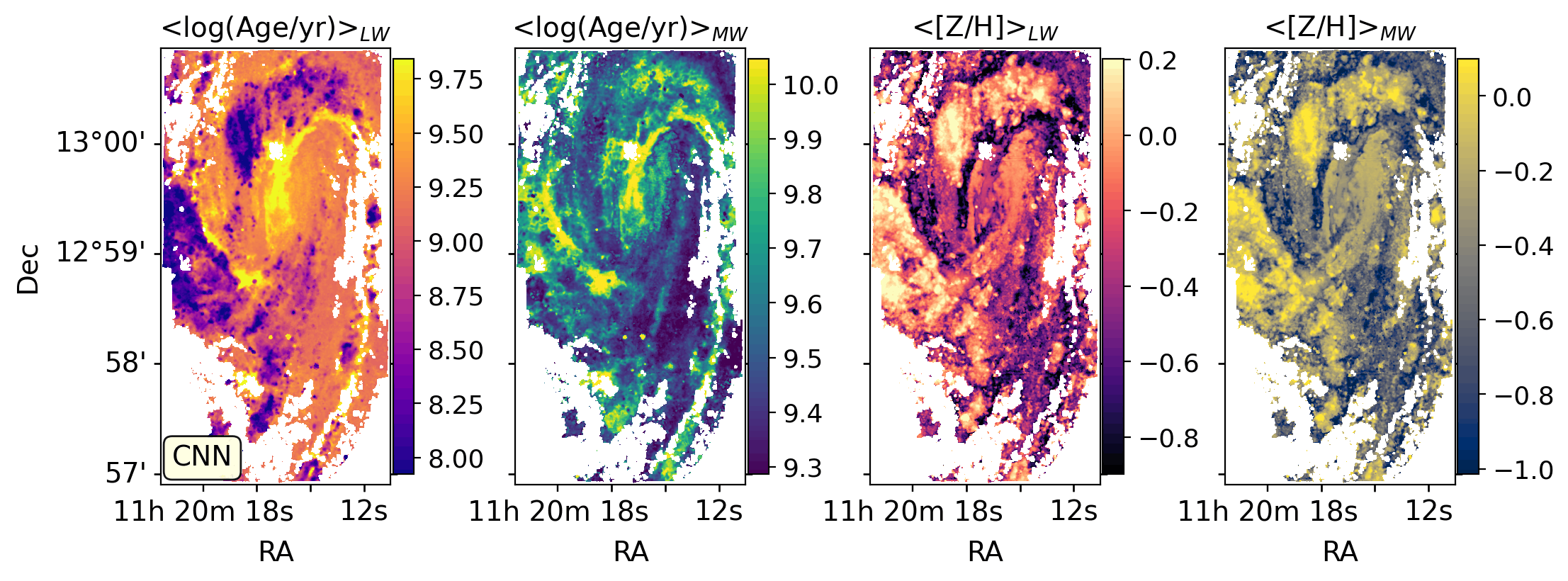}}
  \resizebox{\hsize}{!}{\includegraphics{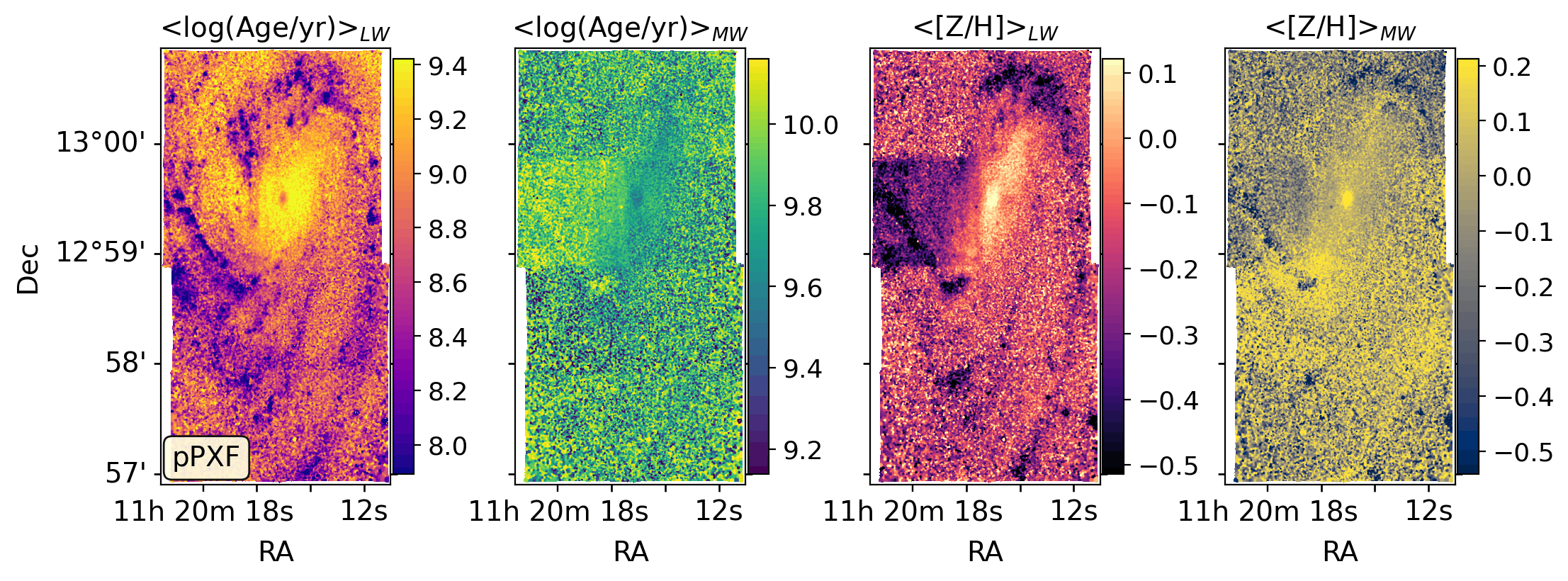}}
  \resizebox{\hsize}{!}{\includegraphics{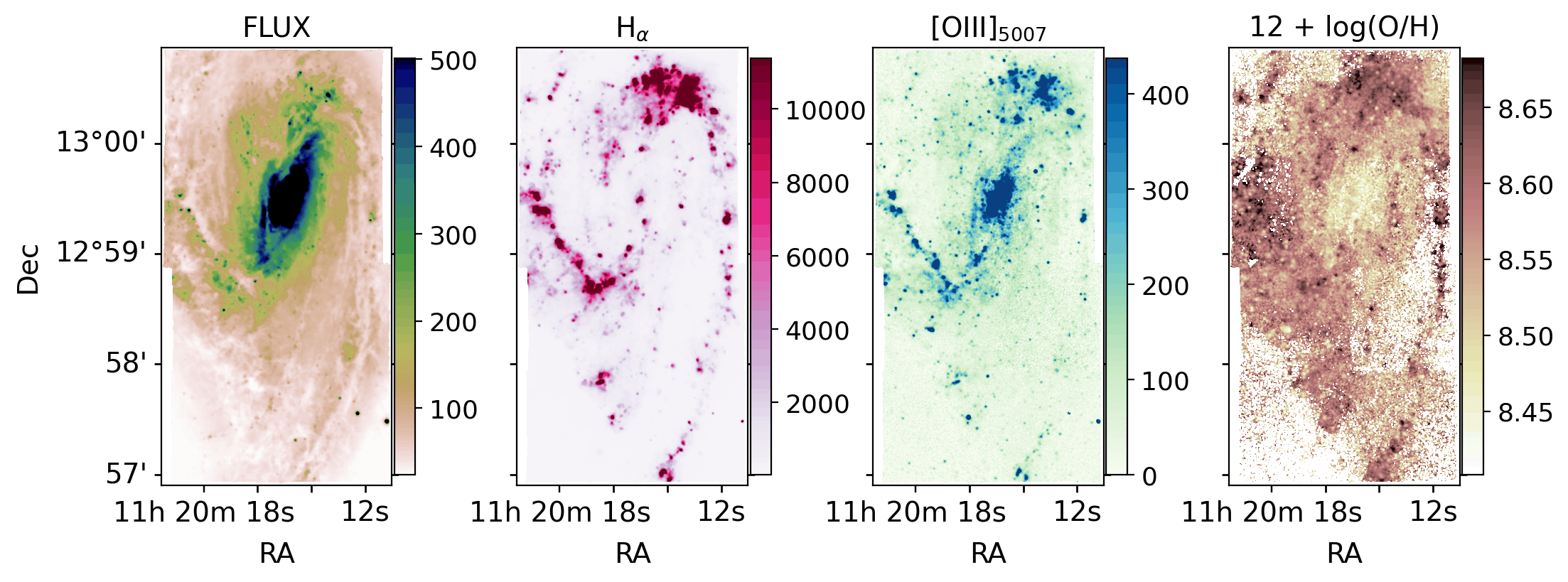}}
  \caption{Top row: maps of the luminosity- and mass-weighted stellar ages and metallicities predicted by the CNN for NGC\,3627. Middle row: equivalent maps obtained with \texttt{pPXF} within the PHANGS-MUSE \texttt{DAP}. Bottom row: maps of the integrated flux, H$\alpha$ and [O\,{\sc iii}]$\lambda5007$ emission, and the metallicity of the ionised gas (see \citealt{Emsellem2022}).}
  \label{fig:maps_cnn}
\end{figure*}

Figure~\ref{fig:maps_cnn} shows the resulting spatially resolved maps of luminosity- and mass-weighted mean ages and metallicities across NGC\,3627. The CNN-derived maps display clear spatial structure, with younger stellar populations tracing the spiral arms and star-forming regions, and older populations dominating the central regions. The maps are smooth and largely free of noise, which would not be expected if the CNN were not able to extract physically meaningful information from real data.

For comparison, we also show the maps obtained with \texttt{pPXF}, as well as maps of the integrated flux and key emission lines (H$\alpha$ and [O\,{\sc iii}]$\lambda5007$). In addition, we include a map of the ionised gas metallicity derived using the O3N2 strong-line indicator, defined as $\log\left(\frac{[\mathrm{OIII}]\lambda5007/\mathrm{H}\beta}{[\mathrm{NII}]\lambda6584/\mathrm{H}\alpha}\right)$, adopting the calibration of \citet{Marino2013}. The emission-line fluxes are corrected for dust attenuation using the Balmer decrement, assuming case-B recombination and a \citet{Cardelli1989} extinction law.

A clear correspondence is observed between regions with low predicted LW ages and enhanced H$\alpha$ emission, confirming that the CNN successfully identifies sites of recent star formation. These regions also tend to exhibit higher metallicities, consistent with expectations from chemical evolution in star-forming environments.

\begin{figure}
  \resizebox{\hsize}{!}{\includegraphics{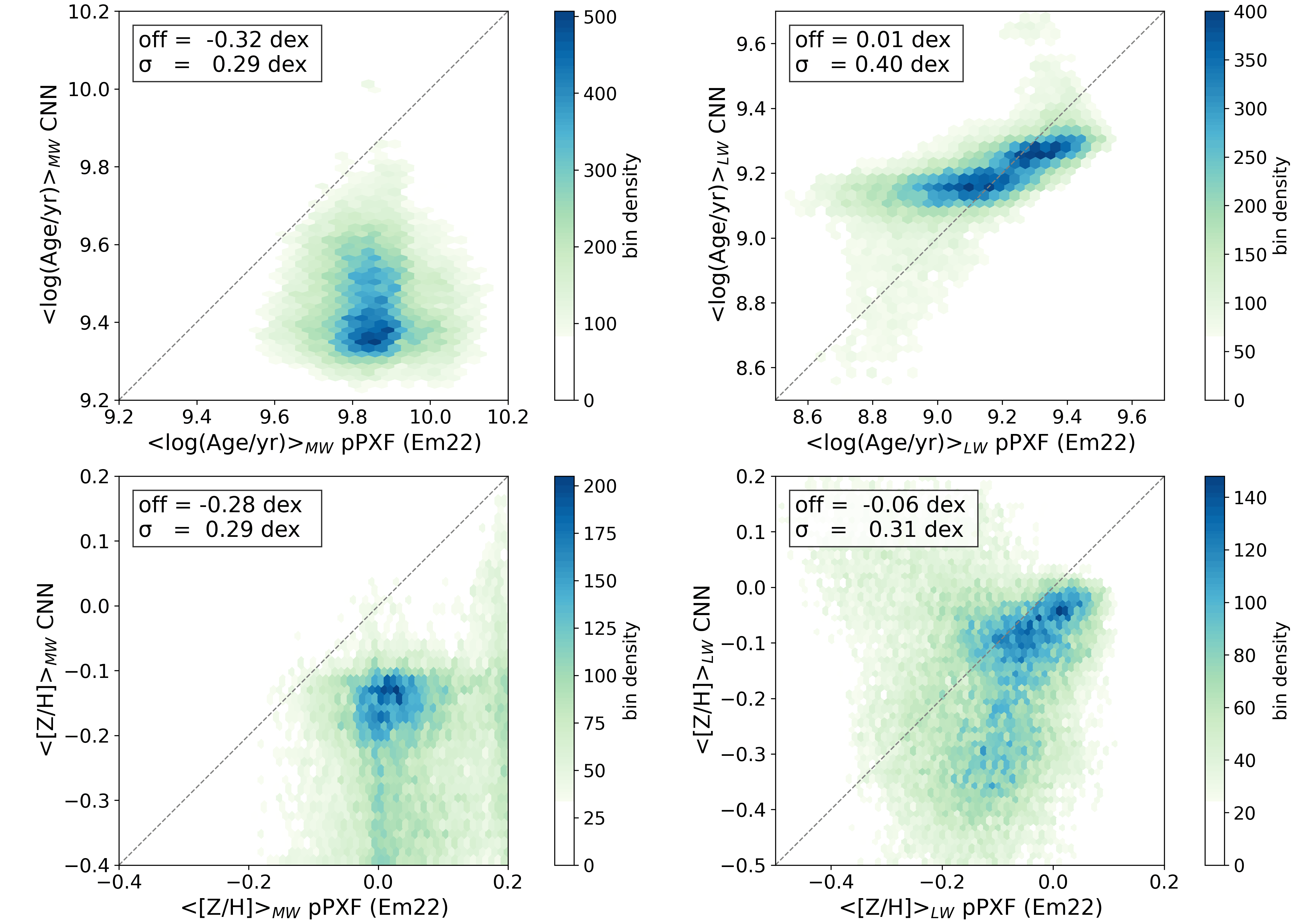}}
  \caption{Comparison of luminosity- and mass-weighted mean stellar ages and metallicities predicted in this work and by \citet{Emsellem2022} for the Voronoi-binned spectra of NGC\,3627. Colours indicate the density of points. The mean offset (CNN $-$ Em22) and the standard deviation of the residuals are reported in the upper-left corner. The one-to-one relation is shown as a dotted line for reference.}
  \label{fig:comparison_ppxf}
\end{figure}

Figure~\ref{fig:comparison_ppxf} presents a quantitative comparison between the CNN and the Em22 estimates for both luminosity- and mass-weighted quantities. The median differences for the luminosity-weighted quantities are relatively small, amounting to 2\% and 12\% for the mean LW age and metallicity, respectively, although with a substantial dispersion (0.4 and 0.31 dex).  In contrast, the mass-weighted quantities exhibit systematic differences of order $\sim50\%$, with the CNN predicting lower values than Em22, corresponding to median offsets of $-0.32$ dex in age and $-0.29$ dex in metallicity.

These results indicate that both methodologies yield broadly consistent estimates for the more recent SFHs, while showing significant discrepancies in the properties of the older stellar populations. This behaviour is expected, as old populations can contribute only weakly to the observed spectra, particularly in star-forming galaxies, where even a small mass fraction of young stars can dominate the emitted light and outshine the flux of older components, making it more difficult to constrain their properties.

Several factors, apart from the analysis method (i.e. \texttt{pPXF} versus the CNN), may be contributing to the observed differences. First, the PHANGS-MUSE \texttt{DAP} analysis relies exclusively on MUSE spectra, whose wavelength coverage does not extend into the bluest regions. These regions are especially sensitive to intermediate and young stellar populations. This limitation makes it more difficult to break the age-metallicity degeneracy in older systems. In contrast, our approach incorporates additional photometric constraints that provide leverage in this regime.  Second, the template library used in the PHANGS-MUSE \texttt{DAP} does not include models older than 14.2~Gyr or with metallicities below [Z/H] $< -0.7$. Additional discrepancies may also arise from differences in the treatment of dust attenuation and nebular emission.

A thorough investigation of these effects is beyond the scope of this work, and a more detailed analysis is deferred to a follow-up study (Galceran et al., in prep.).

\section{Discussion}\label{sec:Discussion}

Recovering the SFH and the stellar metallicity distribution of galaxies remains a challenging inverse problem \citep[see, e.g.][]{Ocvirk2006a, Ocvirk2006b, CidFernandes2023, Leja2019}, where the conversion from observables to the physical solution is highly degenerate.  Full spectral fitting techniques  \citep[][]{CidFernandes2025, Ocvirk2006a, Ocvirk2006b, Cappellari2012, Cappellari2023, Tojeiro2007, Wilkinson2017} have introduced regularisation techniques to minimise these problems, showing they can predict composite populations with different ages and metallicities.  In parallel, stellar population inference has increasingly adopted Bayesian approaches \citep[][]{MartinNavarro2018, Johnson2021},  in which the parameter space is explored using techniques such as Markov Chain Monte Carlo (MCMC). These methods provide a principled way to characterise uncertainties and degeneracies, although they often rely on assumptions about the form of the posterior distribution and can be computationally demanding in high-dimensional parameter spaces.

A limitation of the full spectral fitting techniques is that regions of the spectra highly sensitive to stellar properties have equal weight in the fit compared to spectral regions containing far less information.  In this context, ML approaches have recently emerged as a promising alternative, capable of learning informative spectral features and patterns that may be difficult for conventional full-spectral-fitting techniques to exploit efficiently \citep[see e.g.,][]{Lovell2019, Liew-Cain2021, Iglesias-Navarro2024, Wang2024, Woo2024, Anwar2025, DominguezSanchez2026}. 

As noted by \citet{Walter2026}, a key difference between ML-based and conventional methods lies in the definition of the optimisation target. Whereas traditional spectral-fitting codes minimise the residuals in flux space, CNNs allow minimising the error directly on the physical parameters. The minimisation of parameter-space errors allows the network to identify not only individual features but also higher-order correlations across the spectrum that may carry information relevant to the SFH. An important consequence of this difference in optimisation philosophy is its impact on the classical age-metallicity degeneracy. Several recent studies have shown that CNN-based approaches can substantially mitigate this degeneracy \citep[e.g.][]{Woo2024, DominguezSanchez2026}. By leveraging subtle spectral signatures (such as small variations in the shapes of Balmer lines or weak metal-line blends that do not dominate the global $\chi^2$), ML models can disentangle the effects of age and metallicity more effectively than traditional methods. Moreover, CNNs can, in principle, predict these parameters independently, with age and metallicity uncertainties showing weak correlations \citep{DominguezSanchez2026}, unlike classical fitting approaches where both parameters are strongly coupled through the continuum shape and line-depth trade-off.

The diversity of current methodologies includes transformer architectures optimized for capturing long-range spectral correlations \citep{Anwar2025}, simulation-based inference approaches that combine neural networks with explicit forward modeling to recover full posterior distributions \citep{Iglesias-Navarro2024}, CNNs tailored to extract local spectral features and infer SFHs or integrated stellar population properties \citep[e.g.][]{Lovell2019, DominguezSanchez2026}, and hybrid frameworks that incorporate representation learning, such as autoencoders or latent-space embeddings, to compress spectral information prior to inference \citep[e.g.][]{Liew-Cain2021, Woo2024}. The differences also extend to the training dataset, that includes real data with labels taken from the results of full spectral fitting analysis \cite{Liew-Cain2021}, synthetic data with SFH directly taken from cosmological simulations \citep[e.g.,][]{Lovell2019, Wang2024}, with single stellar populations or a combination of them  \citep[e.g.,]{Anwar2025, DominguezSanchez2026}, or with a similar technique like the one used in this work  \citep{Iglesias-Navarro2024}. Some networks are designed to predict single values of age and metallicities\citep{Woo2024, DominguezSanchez2026},  while others predict vectors quantifying the contribution and the mean metallicity of stars at different ages \citep[][]{Lovell2019, Anwar2025, Iglesias-Navarro2024}. 
 
All these works illustrate the breadth of tools now available.  In this work, we present a new CNN capable of obtaining an accurate SFH and age-metallicity relation from combined spectroscopic and photometric inputs. It is of special interest to compare with those works that, like our model, are able to make predictions of the SFH and metallicities instead of predicting mean values, to investigate the pros and cons of the different configurations and training methods. 

A first natural comparison is with the pioneering work of \citet{Lovell2019}, who trained CNNs on synthetic spectra derived from the EAGLE \citep{Schaye2015-EAGLE} and Illustris \citep{Genel2014-Illustris} cosmological simulations. Their model learns the mapping between rest-frame optical spectra and piece-wise constant SFHs defined in eight logarithmically spaced lookback time bins using a single metallicity. They quantify the differences between the input and predicted SFH using SMAPE, achieving median SMAPE values on the order of 10--12\% for dust-attenuated spectra at SNR = 50. In contrast to our approach, their CNN operates on spectra alone and assumes a single (effectively constant) metallicity for each galaxy. Architecturally, the Lovell et al. network is a relatively shallow 1D CNN with two convolutional layers, max pooling, and fully connected layers performing regression on the SFH bins, optimised with SMAPE as the loss function. This setup is well-suited to capturing the dominant features imprinted by recent star formation. Our model, by comparison, employs a deeper architecture with an attention mechanism, a shared latent space, and a decoder jointly predicting mass fractions and mean metallicities across 16 age bins. It also ingests both spectra and NUV–optical photometry, which improves constraints on young populations. 

A further relevant comparison is provided by the recent work of \citet{Woo2024}, who also employs a deep-learning CNN (\texttt{STARNET}; \citet{Fabbro2018}) to infer stellar population properties (mass-weighted ages, metallicities, and $M_\star L_r$ ratios) and colour excess from galaxy spectra. Their network is trained on synthetic spectra built from SFHs from the \texttt{IllustrisTNG100-1} simulation and adapted to mimic the Sloan Digital Sky Survey \citep[SDSS;][]{York2000}. The authors include realistic noise that incorporates sky residuals and emission lines taken from real, randomly chosen SDSS spectra that minimise the domain shift. 

Their model can predict the mass-weighted ages and metallicities with very small biases of 0.0032 and 0.017 dex, respectively, with scatters of 0.077 and 0.081 dex, respectively. In comparison, our predictions show a lower bias in the recovery of the mass-weighted age and metallicity of $-0.0007$ and 0.0003 dex, respectively, with dispersions of 0.117 and 0.042 dex. Overall, the values are comparable. However, there are important methodological differences between the two approaches. The model presented by \citet{Woo2024} is primarily designed to predict integrated summary quantities (e.g., mean ages and metallicities), rather than reconstructing the full star formation and chemical enrichment histories. As a result, their network architecture and training objective are optimised for capturing global spectral trends, rather than resolving the temporal structure of the SFH. In contrast, our model explicitly targets the recovery of the full SFH and age-metallicity relation across multiple age bins, which requires a more expressive architecture and a training strategy tailored to high-dimensional outputs. 

Furthermore, our method incorporates emission lines self-consistently, linking their presence and strength to the population of young stars capable of ionising the gas. In contrast, \citet{Woo2024} includes emission lines drawn from real spectra, which may result in more realistic line ratios and profiles. However, these emission features are added randomly, without any physical correlation with the underlying stellar population or ionising radiation field.

This work also presents a detailed comparison with several full spectral fitting methods. Consistent with our findings, the authors report that ML models recover stellar population parameters with reduced bias and weaker age-metallicity degeneracies. They further show that, while emission lines significantly affect the results of traditional full spectral fitting codes, they have minimal impact on the CNN predictions. We argue that this behaviour likely stems from the random inclusion of emission lines, which are not physically linked to the stellar populations that ionise the gas. As a result, their model cannot extract meaningful information from these features, in contrast to our approach, where emission lines carry physically relevant information.

Another comparison point is provided by the transformer-based architecture of {\tt GalProTe} from \citet{Anwar2025}. That model uses four parallel attention-based encoders with multi-scale kernels to process optical spectra and predicts the weights of a dense grid of stellar populations sampling 12 metallicities and 53 ages. The inclusion of self-attention and multi-scale convolution allows {\tt GalProTe} to learn cross-dependencies between spectral regions and to capture correlations between age- and metallicity-sensitive features that are widely separated in wavelength.  The model is trained with synthetic spectra tailored to characteristics of the PHANGS-MUSE survey, with SFH that are linear combinations (from one to five) of single stellar populations, randomly selected from the 53$\times$12 age-metallicity grid of templates with different levels of extinctions and noise. The authors include Gaussian noise, but they also test the effect of adding very high noise in regions of the spectra affected by strong sky residuals. The self-attention mechanisms help the network learn cross-dependencies across spectral regions and therefore help to break degeneracies between parameters, achieving small biases in mean stellar ages ($\sim$0.02~dex) with moderate scatter ($\sim$0.25~dex). However, the metallicity estimates are notably less accurate, likely reflecting the relatively simple SFH parametrisations used in its synthetic training set, with metallicities that can change rapidly between adjacent age bins. 

A conceptually different approach is presented by \citet{Iglesias-Navarro2024}, who adopts a simulation-based inference (SBI) framework. Their method compresses spectra using a convolutional autoencoder and then applies a normalising-flow density estimator to recover full posterior distributions for a set of cumulative stellar-mass percentiles and the metallicity. This approach provides formally Bayesian uncertainty quantification and captures the multi-modal posteriors that naturally arise in SFH inference. However, because the autoencoder is trained to optimise spectral reconstruction rather than direct recovery of physical parameters, its latent space may encode information that is not maximally informative for SFH inference. In contrast, our CNN is trained end-to-end on the desired physical outputs (SFH and age-metallicity relation), ensuring that the latent representation is explicitly tuned to the astrophysical parameter space rather than to spectral fidelity.

The two approaches thus offer complementary strengths. SBI provides rigorous posterior estimation under controlled domain assumptions but is more sensitive to domain shifts between simulated training data and real spectra. Our CNN lacks a full probabilistic posterior and therefore cannot characterise multi-modal solutions in the same way; however, it achieves substantially higher accuracy in detailed SFH reconstruction, integrates photometric and spectroscopic information naturally, mitigates age-metallicity degeneracy more effectively, and performs inference at significantly lower computational cost, which makes it particularly interesting for application to large spectroscopic surveys. Furthermore, our network parameterises the SFH with the SFR(t)  vector rather than mass percentiles. This enables our network to better discriminate the variations of the SFH at young ages.  For many studies that target mainly intermediate or older populations, a percentile-based parametrisation will suffice.

Despite these successes, the present method also has limitations. The requirement to shift spectra to rest-frame (z = 0) and broaden them to a fixed velocity dispersion ($\sigma$ = 150~km~s$^{-1}$ introduces a preprocessing step that partly offsets the rapid inference speed of the NN. More fundamentally, supervised ML models depend on the availability of labelled training data. Constructing such training sets requires explicit assumptions about extinction curves, the escape fraction of ionising photons, the initial mass function, and other astrophysical ingredients. Any mismatch between these assumptions and the true underlying physics may introduce biases in the predictions. This issue is not unique to ML (errors in stellar-population models propagate through all inference frameworks), but the impact is particularly direct for supervised systems. Previous studies \citep[e.g.][]{Kang2023, Ksoll2024} have emphasised the importance of addressing such model-dependence and have suggested that transfer-learning techniques \citep{Farahani2021} may offer a way forward by adapting networks trained on synthetic data to better match real observations.

\section{Summary}\label{sec:summary}
In this work, we have presented a convolutional neural network (CNN) capable of recovering detailed star formation histories (SFHs) and age-metallicity relations directly from combined spectroscopic and photometric inputs. The model was trained on synthetic observations generated from a wide range of SFH and chemical-enrichment histories. Dust attenuation was modelled following the two-component prescription of \citet{Charlot2000}, with independently sampled optical depths for birth clouds and the diffuse interstellar medium. Emission lines associated with young stellar populations were added using the MAPPINGS-III photoionisation models, and Poisson noise was incorporated at varying signal-to-noise ratios.

When tested against synthetic data, the CNN attains higher accuracy and precision than {\tt pPXF} (v9.4.2), one of the most widely used full spectral-fitting codes. It produces lower biases and reduced scatter in the predicted luminosity- and mass-weighted ages and metallicities, while also delivering predictions at sub-millisecond speeds. Similar gains have been reported by recent machine-learning studies focused on mean stellar-population parameters \citep[e.g.][]{Liew-Cain2021, Wang2024, Woo2024, DominguezSanchez2026} and full SFHs \citep{Lovell2019, Anwar2025, Iglesias-Navarro2024}. Our model extends these efforts by predicting the full metallicity evolution and exploiting a combination of spectroscopic and photometric constraints.

The good performance of the model can be attributed to the fact that the neural network is optimised to recover physically meaningful SFHs rather than to minimise spectral residuals alone. Traditional fitting methods based on explicit residual minimisation---such as $\chi^{2}$ optimisation---can sometimes achieve deceptively small variances by adjusting broad continuum shapes while failing to reproduce age- and metallicity-sensitive spectral features. In contrast, our CNN is trained directly on the underlying SFHs and enrichment histories, favouring solutions that more accurately reflect the physical parameters even when they do not correspond to the smallest possible spectral residuals. Our analysis further shows that including photometric constraints, particularly in the UV--optical range, plays a crucial role in breaking degeneracies and enhancing age sensitivity.

Our methodology is specifically designed to recover SFHs and age--metallicity relations from PHANGS-MUSE and PHANGS-HST data, and the network architecture reflects the characteristics of these observations. The resulting model is therefore not intended as a universal SFH inference tool that can be applied unchanged to arbitrary datasets. However, the approach itself is readily transferable to other spectral ranges or to different combinations of broad- and narrow-band photometry, provided that the architecture is adjusted and the network retrained accordingly. In this sense, the main contribution of this work is not only the trained model presented here, but also a reproducible framework for generating synthetic training data, defining an appropriate architecture, and validating the inference. Although developed in the context of PHANGS, this strategy is more broadly applicable and illustrates how convolutional architectures can be tailored to mixed spectroscopic and photometric datasets while preserving interpretability and training stability.

The application of the CNN to real PHANGS-MUSE data demonstrates its ability to recover physically meaningful stellar population properties from observed spectra. The spatial distributions of age and metallicity are consistent with the known structure and star formation activity of NGC\,3627, and the expected abundance gradients. A follow-up study (Galceran et al., in prep.) will extend this application to 19 galaxies from the PHANGS-MUSE and PHANGS-HST surveys, enabling an exhaustive comparison of the CNN performance in real galaxies.

Importantly, the CNN offers significant advantages in terms of computational efficiency and robustness to noise, enabling the analysis of large spectroscopic datasets at a fraction of the computational cost of traditional fitting methods. Moreover, the smoother spatial behaviour of the inferred quantities suggests that the CNN is less affected by local degeneracies and noise-induced fluctuations.

These results contribute to a rapidly growing body of evidence demonstrating the potential of deep learning approaches as a complementary tool to classical full spectral fitting techniques for the analysis of spatially resolved stellar populations in nearby galaxies.

\begin{acknowledgements}
We thank the anonymous referee for a careful and thorough revision of the manuscript and for constructive comments that have significantly improved the final presentation of this paper.
We acknowledge financial support by the Spanish Ministry of Science and Innovation through the research grants PID2019-107427-GB-31,  funded by MICIU/AEI/10.13039/501100011033 and the European Union/FEDER, PID2022-138855NB-C31,  funded by  MICIU/AEI/10.13039/501100011033 and PRE2020-094548, funded by MICIU/AEI/10.13039/501100011033 and ESF Investing in your future.
MB acknowledges support from the ANID BASAL project FB210003. This work was supported by the French government through the France 2030 investment plan managed by the National Research Agency (ANR), as part of the Initiative of Excellence of Université Côte d’Azur under reference number ANR-15-IDEX-01.
FP acknowledges support from the Horizon Europe research and innovation programme under the Maria Skłodowska-Curie grant “TraNSLate” No 101108180, and from the Agencia Estatal de Investigación del Ministerio de Ciencia e Innovación (MCIN/AEI/10.13039/501100011033) under grant (PID2021-128131NB-I00) and the European Regional Development Fund (ERDF) ``A way of making Europe''. 
RSK acknowledges financial support from the ERC via Synergy Grant ``ECOGAL'' (project ID 855130),  from the German Excellence Strategy via the Heidelberg Cluster ``STRUCTURES'' (EXC 2181 - 390900948), from the German Ministry for Economic Affairs and Climate Action in project ``MAINN'' (funding ID 50OO2206), and from DFG and ANR for project ``STARCLUSTERS'' (funding ID KL 1358/22-1). 
This work was carried out as part of the PHANGS collaboration.
Based on observations taken as part of the PHANGS-MUSE large program (Emsellem et al. 2021).
Based on data products created from observations collected at the European Organisation for
Astronomical Research in the Southern Hemisphere under ESO programme(s) 1100.B-0651, 095.C-0473,
and 094.C-0623 (PHANGS-MUSE; PI Schinnerer), as well as 094.B-0321 (MAGNUM; PI Marconi),
099.B-0242, 0100.B-0116, 098.B-0551 (MAD; PI Carollo) and 097.B-0640 (TIMER; PI Gadotti). This
research has made use of the services of the ESO Science Archive Facility.
This research has made use of the services of the ESO Science Archive Facility.
Science data products from the ESO archive may be distributed by third parties and disseminated via
other services, according to the terms of the Creative Commons Attribution 4.0 International license.
Credit to the ESO origin of the data must be acknowledged, and the file headers preserved.
\end{acknowledgements}

\bibpunct{(}{)}{;}{a}{}{,}
\bibliographystyle{aa}
\bibliography{bibtex}

\appendix
\section{Stability of the loss function to the training size}
\label{Appendix:Loss}
\begin{figure}[ht]
  \resizebox{\hsize}{!}{\includegraphics{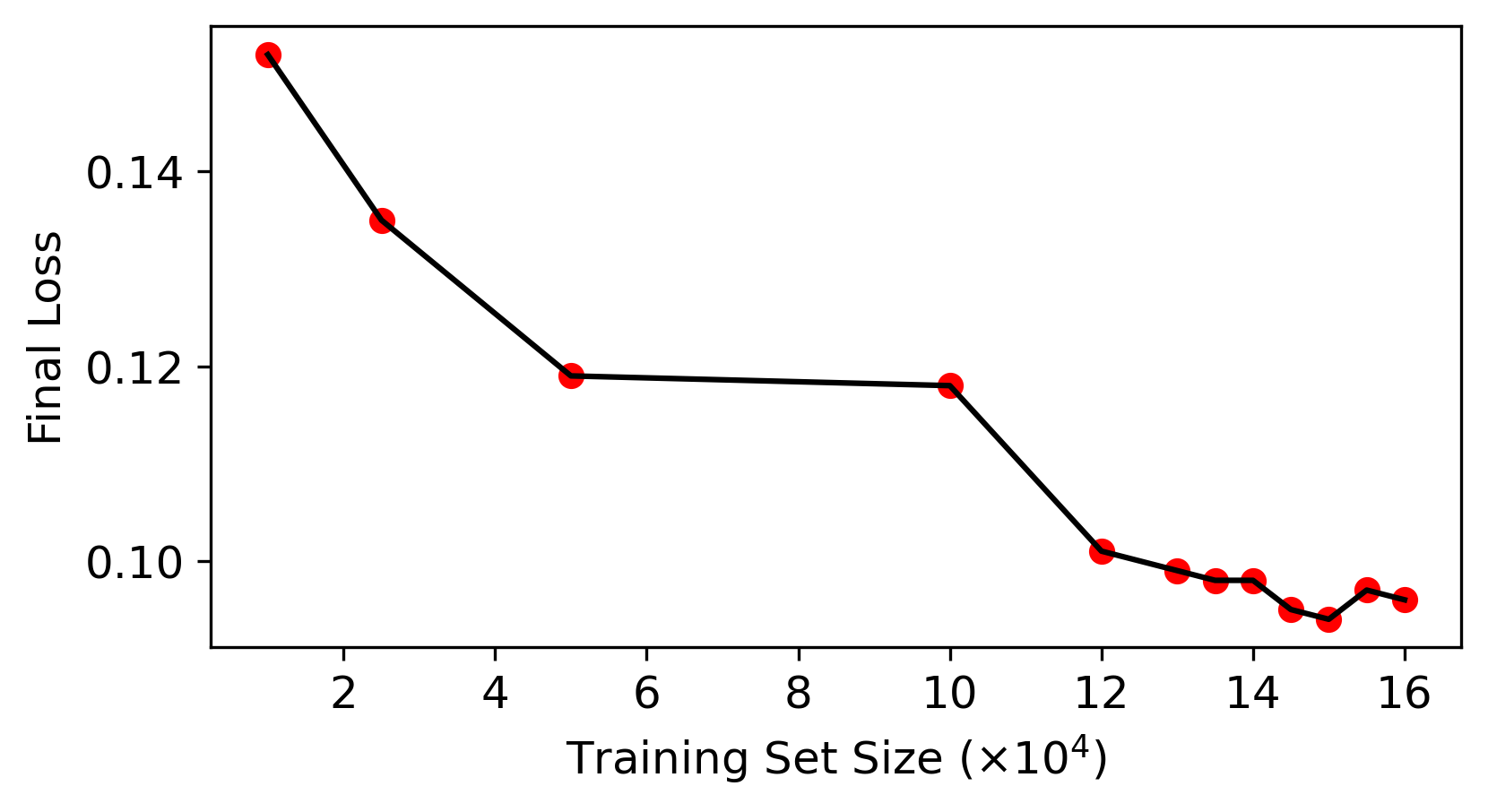}}
\caption{Final converged validation loss as a function of the number of training samples used.}
\label{fig:loss_function}
\end{figure}

Figure~\ref{fig:loss_function} shows the final value of the loss function reached by our CNN when trained with random samples of different sizes. The final loss decreases monotonically with increasing dataset size up to $\approx$125\,000, reaching a plateau beyond this point, except for the small fluctuations expected in the training process. This behaviour indicates that the adopted training-set size is sufficient to constrain the model and mitigates concerns about overfitting driven by an undersampled training distribution.

\section{Age-metallicity correlations}
\label{sec:Age_Z_correlations}
\begin{figure}[ht]
  \resizebox{\hsize}{!}{\includegraphics{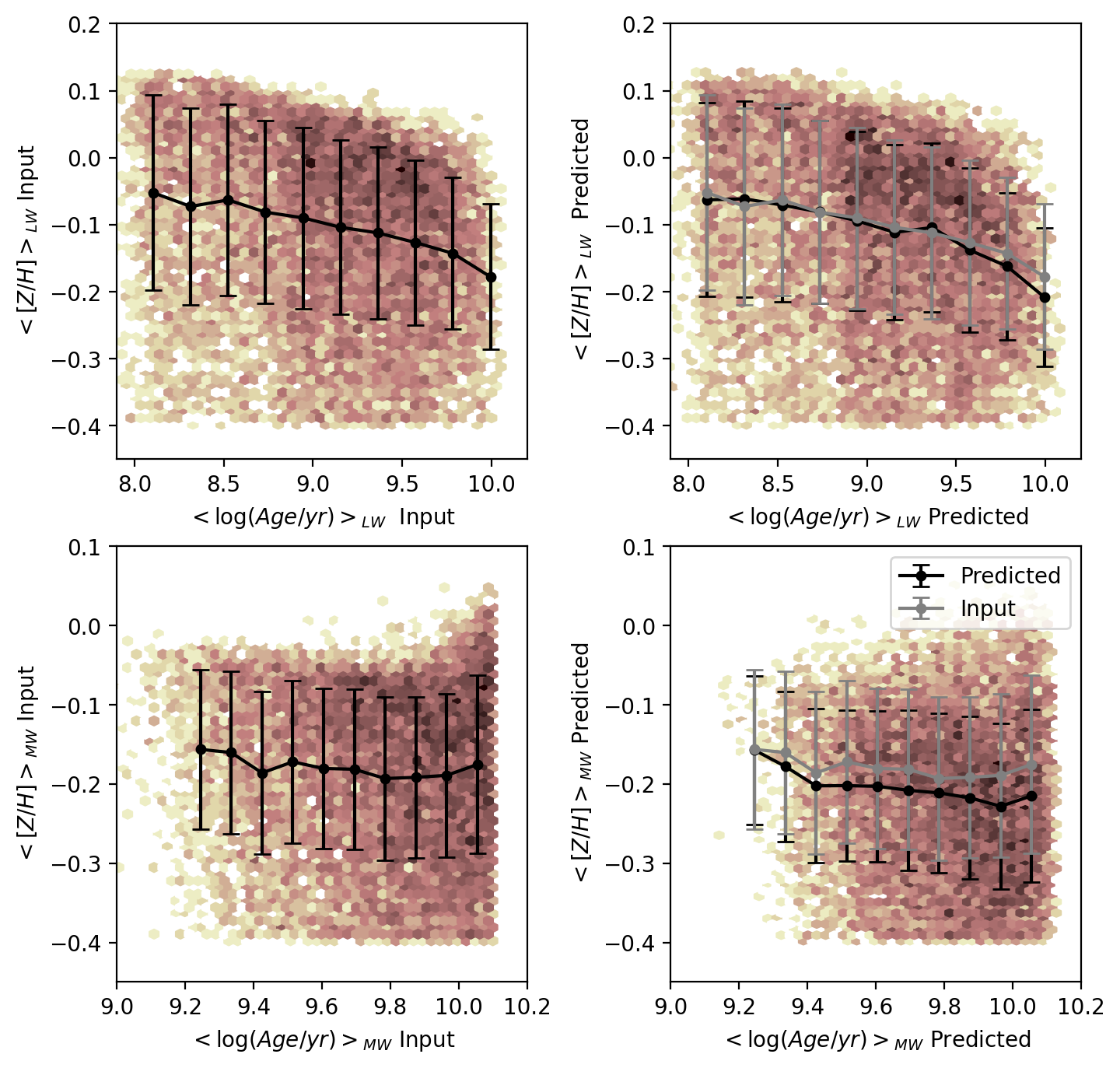}}
\caption{Age-metallicity relation for the input (left column) and predicted (right column) data. The panels show the relation between the mean $<\log\left(\text{Age}(yr)\right)>$ and mean $<[Z/H]>$ weighted with luminosity (top row) and mass (bottom row). The colours represent the density of points in hexbins.}
\label{fig:Age_Z_correlations}
\end{figure}

Figure \ref{fig:Age_Z_correlations} shows the mean light-weighted (top row) and mass-weighted (bottom row) metallicities as a function of age for the synthetic test set (unseen during training). The left panels show the ground-truth values, and the right panels show the CNN predictions. Each panel displays a logarithmic hexbin density map. The predicted relations closely follow the input ones, indicating that the network preserves the underlying trends present in the synthetic data and that these relations are not washed out by dust, noise, or the range of chemical enrichment parameters. 

\section{UMAP representations}
\label{app:UMAP_representation}
\begin{figure}[htb]
        \centering
        \includegraphics[width=\linewidth]{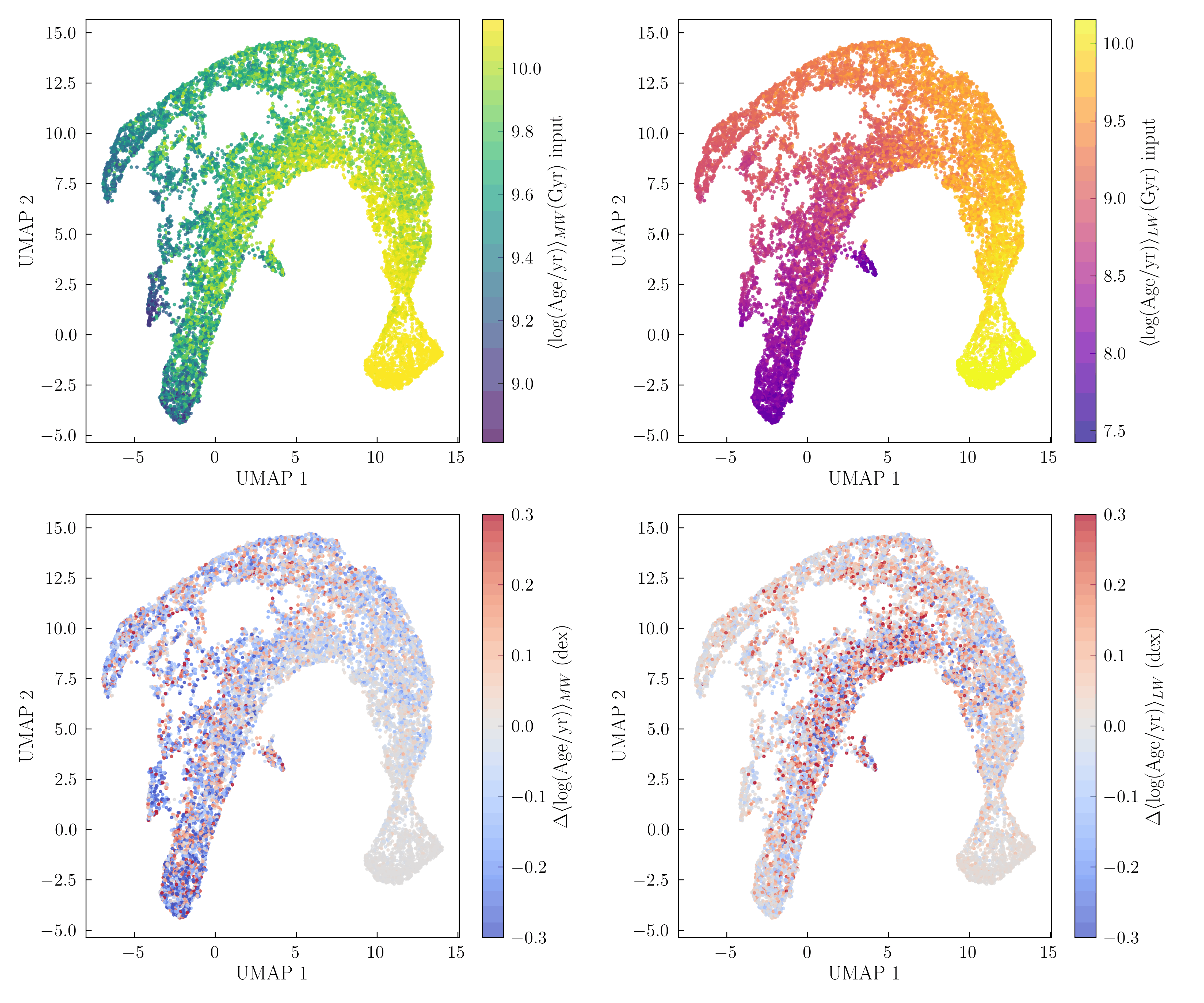}
        \caption{UMAP projection of the 40-dimensional latent space in the test set and corresponding CNN predictions. Each panel shows the same 2-D embedding, coloured by either the CNN's prediction (top row) or the relative error (bottom row) for the $\langle \log(\mathrm{Age}/\mathrm{yr}) \rangle_{\rm MW}$ (left columns) and $\langle \log(\mathrm{Age}/\mathrm{yr}) \rangle_{\rm LW}$ (right columns).}
        \label{fig:umap1}
    \end{figure}

\begin{figure}[htb]
        \centering
        \includegraphics[width=\linewidth]{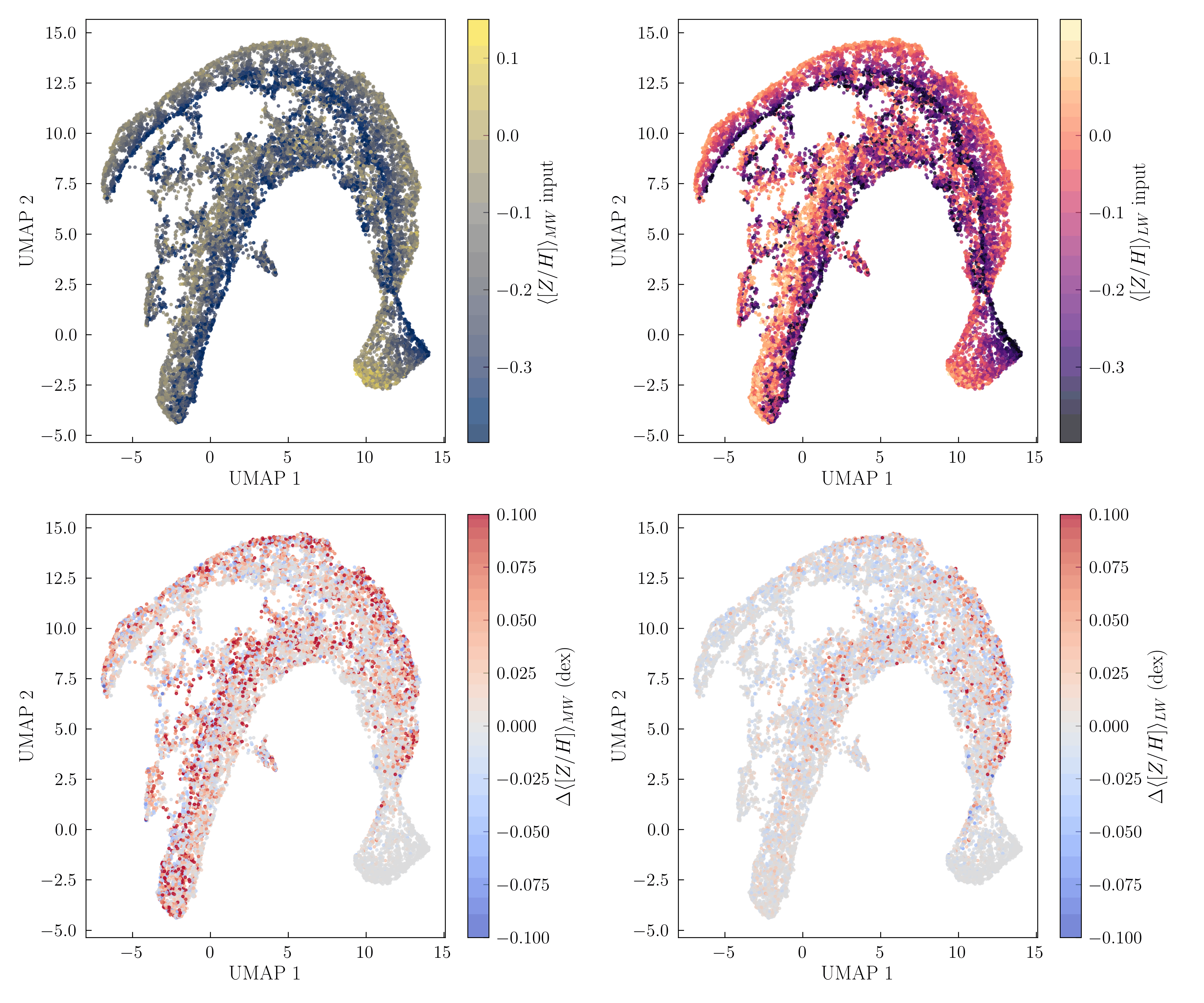}
        \caption{UMAP projection of the 40-dimensional latent space in the test set and corresponding CNN predictions. Each panel shows the same 2-D embedding, coloured by either the CNN's prediction (top row) or the relative error (bottom row) for the \zmw{} (left columns) and \zlw{} (right columns).}
        \label{fig:umap2}
    \end{figure}

\begin{figure}[htb]
        \centering
        \includegraphics[width=\linewidth]{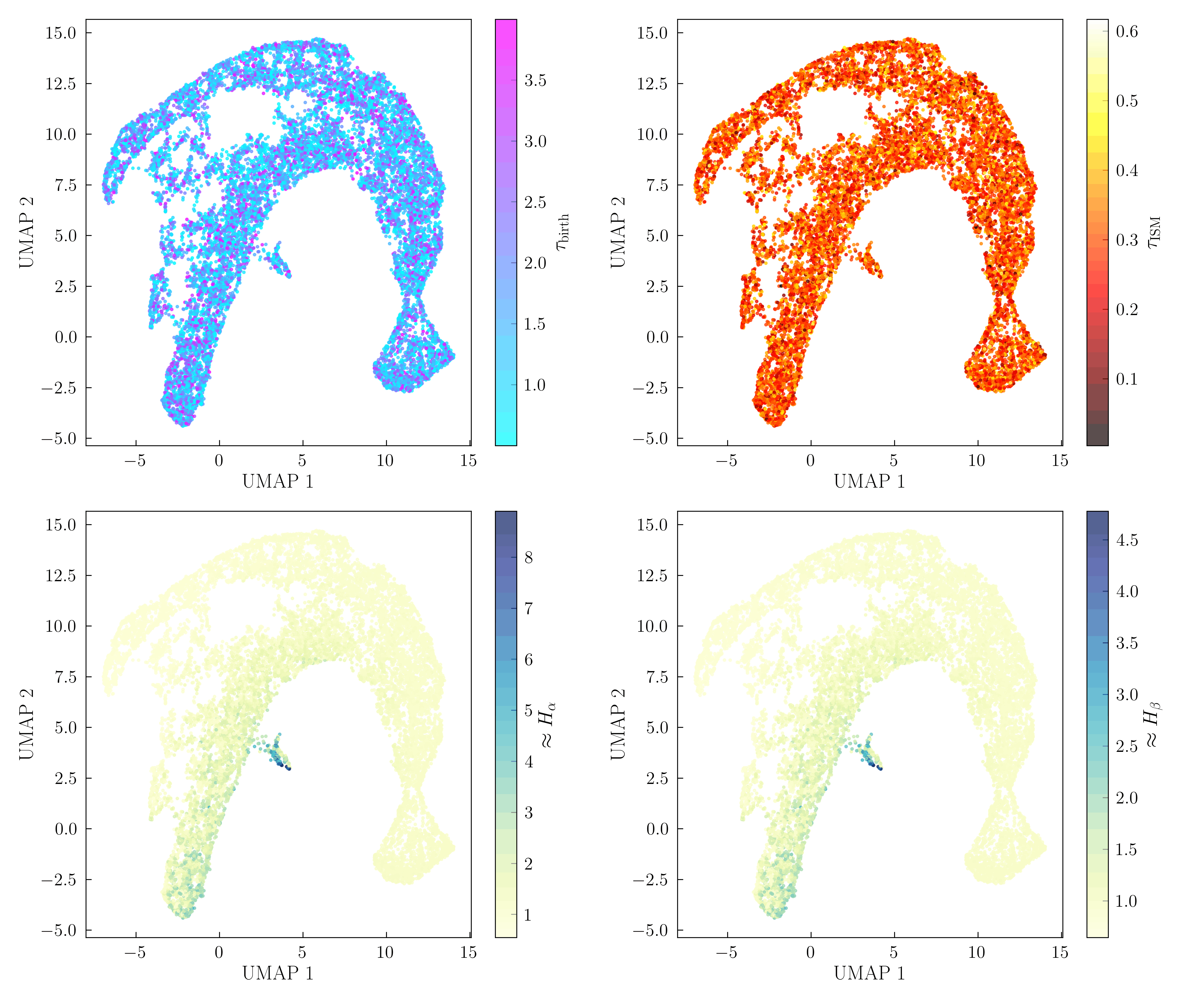}
        \caption{UMAP projection of the 40-dimensional latent space in the test set and corresponding CNN predictions. Each panel shows the same 2-D embedding, coloured (from top to bottom, and left to right) by the extinction values $\tau_\text{birth}$ and $\tau_\text{ISM}$, and by the relative strength of $H_\alpha$ and $H_\beta$ Balmer lines.}
        \label{fig:umap3}
    \end{figure}

To assess the structure of the test dataset and the behaviour of the CNN across different regions of parameter space, we projected the 40-dimensional latent space onto a 2-D manifold using Uniform Manifold Approximation and Projection \citep[UMAP;][]{McInnes2018}. Figures~\ref{fig:umap1} and \ref{fig:umap2} present the resulting UMAP projection with the colour representing the $\langle \log(\mathrm{Age}/\mathrm{yr}) \rangle$ and $\langle \log([Z/H])\rangle$, respectively. The bottom row represents the difference in the input and predicted values (in dex). The resulting embedding exhibits a smooth and continuous geometry, with no disconnected clusters, indicating that the latent space forms a coherent family rather than a mixture of isolated modes. Physical quantities such as mass- and luminosity-weighted age and metallicity display clean, monotonic gradients across the manifold, demonstrating that the UMAP projection organises the latent space according to physically meaningful variations. Importantly, the residual maps show no coherent regions of systematic over- or under-prediction, indicating that the CNN performs consistently across the full latent space manifold. In Fig.~\ref{fig:umap3}, we present the dust extinction of the data and the strength of the $H_\alpha$ and $H_\beta$ Balmer lines. The extinction is homogenously distributed along the space and does not drive the topology, confirming that the extinction plays a secondary role in defining the intrinsic structure of the dataset. The only slightly isolated clustered regions of the UMAP projection are located in the small section with high Balmer line measurements.
Together, these diagnostics demonstrate that the test data populate a well-behaved physical manifold and that the network learns this structure robustly, without topology-dependent biases.

\section{Performance of the CNN to data with realistic noise}\label{app:realistic_noise}

\begin{figure}[htb]
        \centering
        \includegraphics[width=\linewidth,trim=4.25cm 1cm 3.5cm 2.7cm,clip]{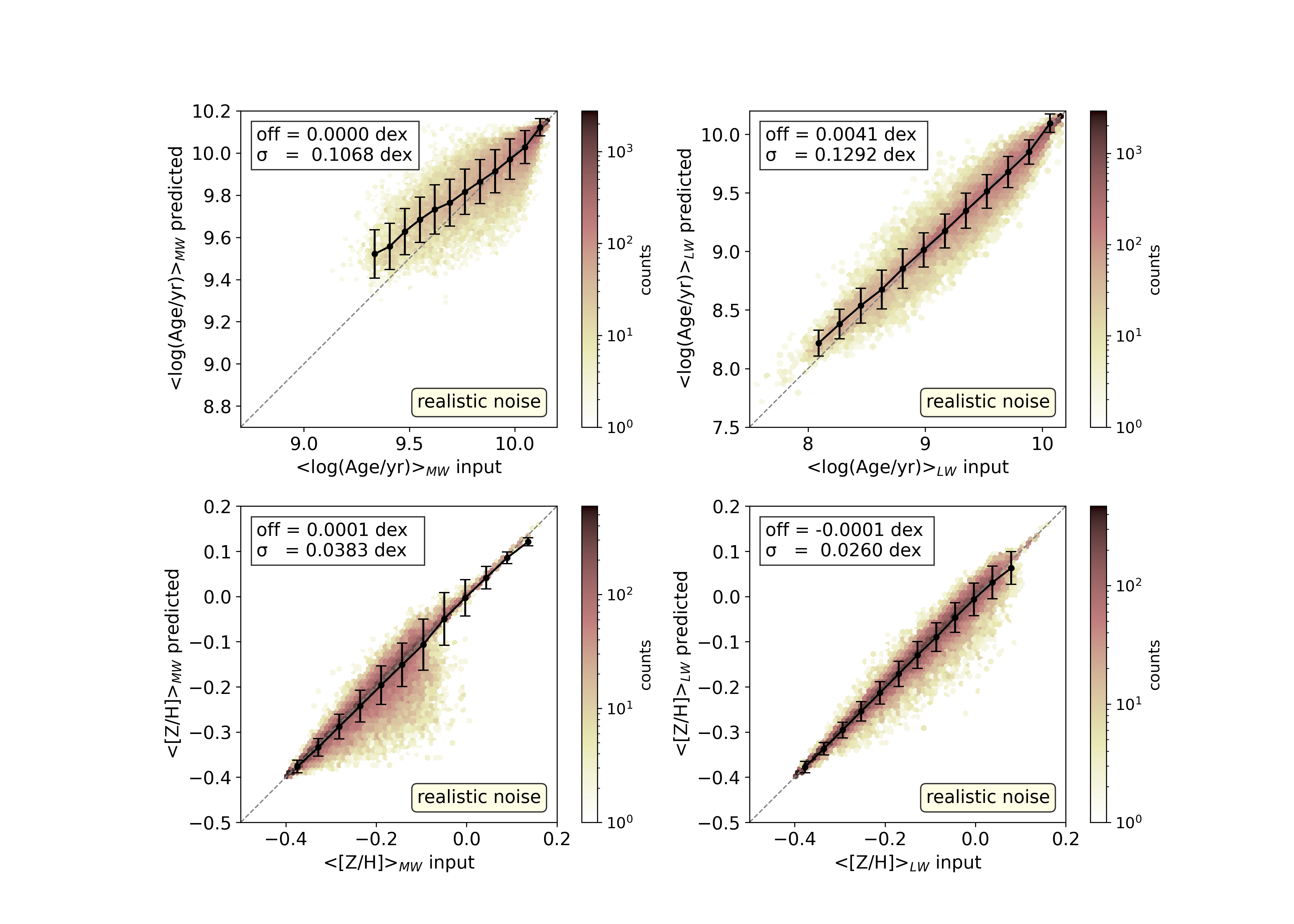}
        \caption{Comparison between the ground truth and the predicted mean mass- and luminosity-weighted ages (top row) and metallicities (bottom row) for test data altered with "realistic" noise (see text). The insets display the mean logarithmic differences between the input and predicted values ($\langle \log(\mathrm{Age/yr})_{\rm input} - \log(\mathrm{Age/yr})_{\rm predicted} \rangle$ and $\langle [Z/H]_{\rm input} - [Z/H]_{\rm output} \rangle$), together with their standard deviations. Each hexbin is coloured according to its point density, as indicated by the colour bar. Black symbols denote the median predicted values within bins of the corresponding input quantities.}
        \label{fig:Comparison_means_RealisticNoise}
    \end{figure}

To assess the robustness of the model under more realistic observational conditions, we perform an additional test in which synthetic spectra are perturbed using noise derived from real data. Specifically, we construct noise from the RMS of the residuals obtained by fitting PHANGS-MUSE spectra with \texttt{pPXF} and alter each pixel of the spectra randomly with Gaussian distributions of width given by these residuals. This approach captures not only photon noise but also systematic features present in real observations, such as residuals from strong sky emission lines and other instrumental or reduction-related imperfections. Incorporating this type of noise provides a more faithful representation of the data domain and allows us to evaluate the impact of potential domain shifts between the synthetic training set and real observations. Figure~\ref{fig:Comparison_means_RealisticNoise} compares the mean MW- and LW-ages and metallicities recovered by the model under these conditions with the ground truth values. The bias and the dispersion in the predicted values do not show any statistically meaningful difference with those obtained with the data with Poisson noise (see Fig.~\ref{fig:comparison_agemw_hexbin}), demonstrating that our model is robust to those imperfections of the spectra not correlated with the physical outputs of the CNN.

\section{Spectral segments and transmission responses used in the ablation tests}

\begin{figure}
\centering
\resizebox{\hsize}{!}{\includegraphics{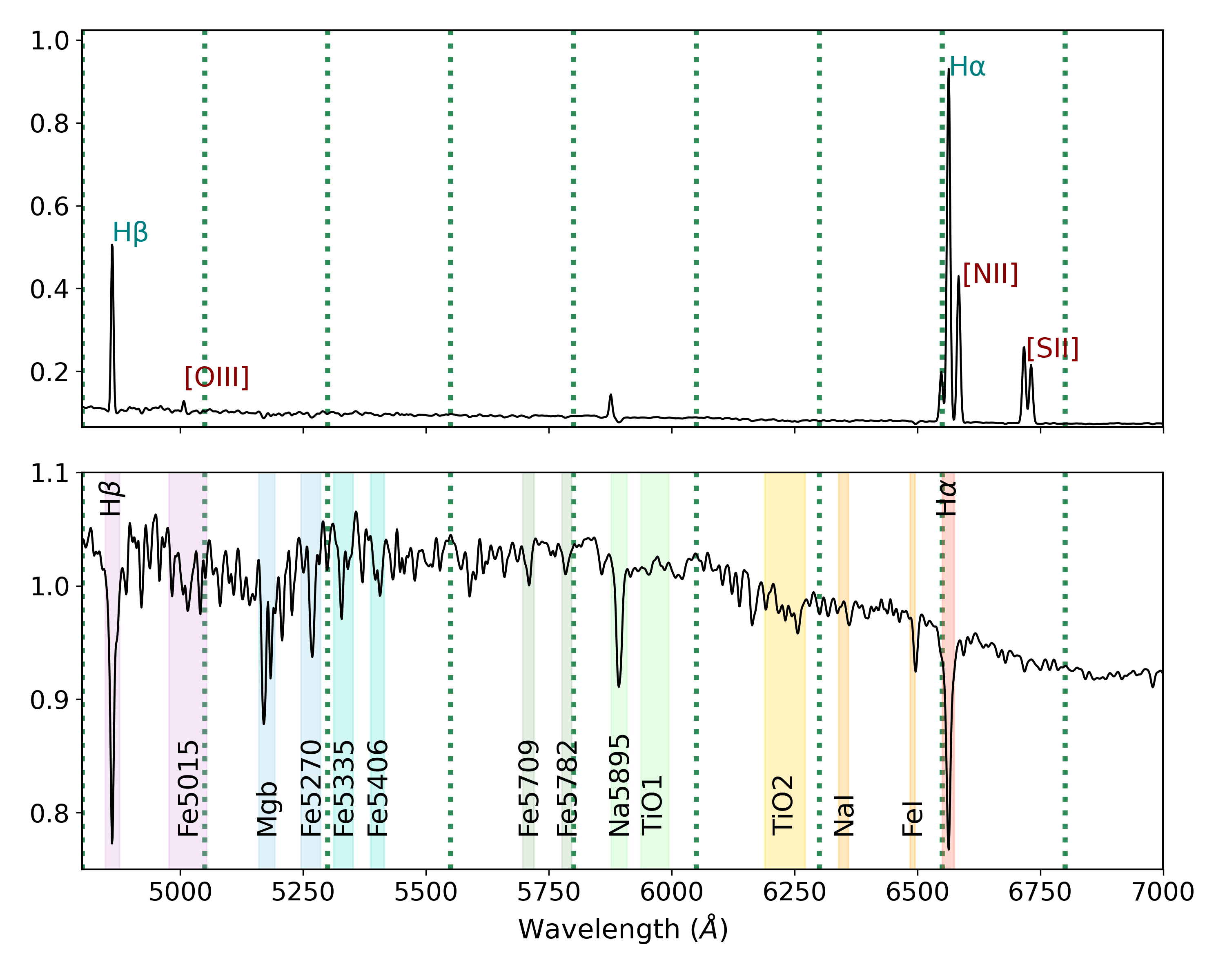}}
\caption{Main emission (top panel) and absorption (bottom panel) features in the spectral range 4780--7000~\AA. The central passbands of the Lick/IDS indices \citep{Worthey1994-lick-indices} are shown as shaded regions. The dotted vertical lines mark the nine spectral segments used in the ablation tests.}
\label{fig:spectra_lines}
\end{figure}

\begin{figure}
\centering
\resizebox{\hsize}{!}{\includegraphics{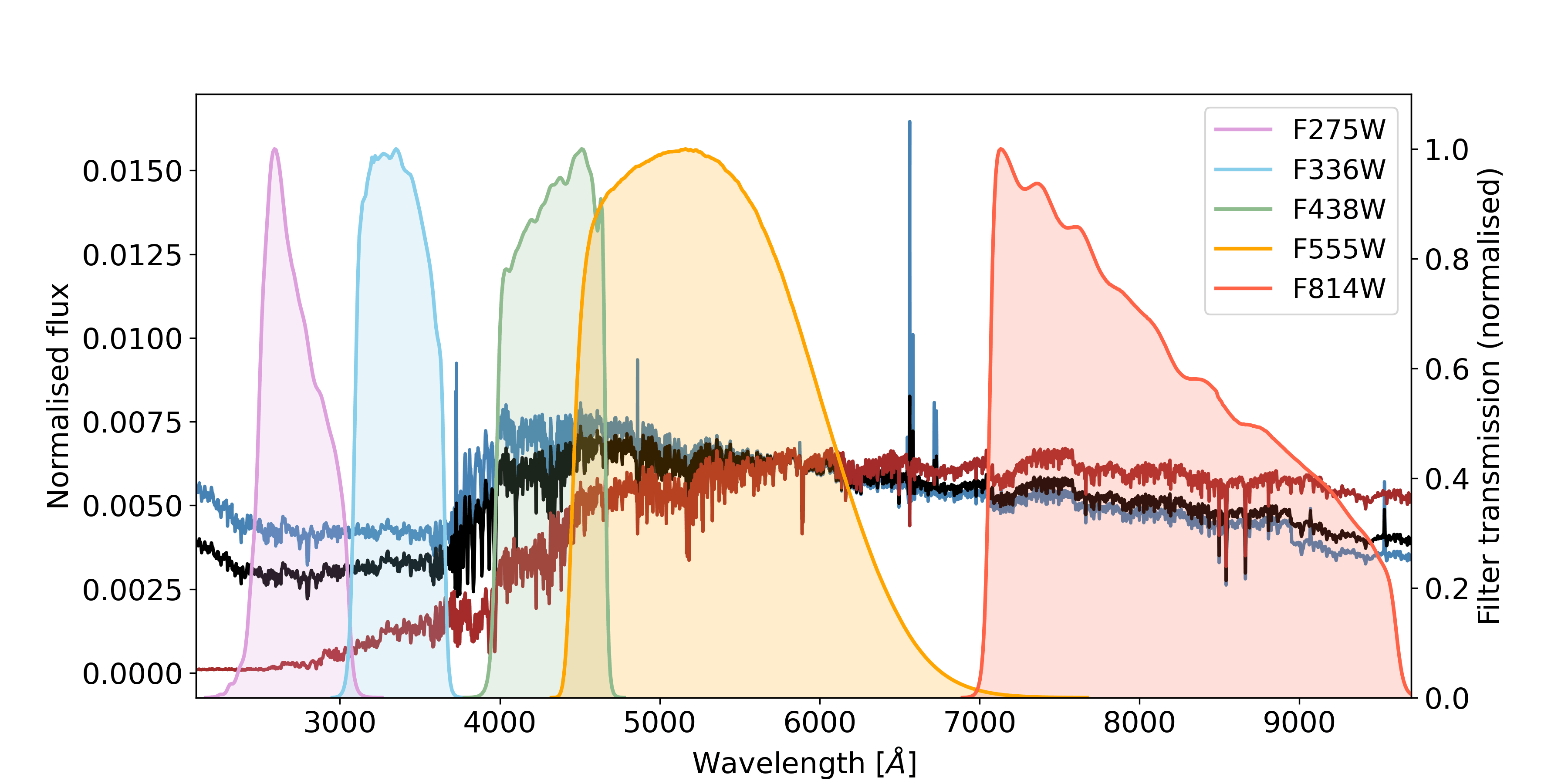}}
\caption{Transmission curves of the HST/WFC3 filters used in this work (F275W, F336W, F438W, F555W, and F814W), overlaid on three example spectra from the test sample. The spectra have similar mean metallicity ($\langle [Z/H] \rangle_{\rm LW} = 0.17$) but different light-weighted ages: $\langle \log(\mathrm{Age}/\mathrm{yr}) \rangle_{\rm LW} = 8.9$ (blue), $9.6$ (black), and $10.1$ (red). This comparison illustrates how the relative flux sampled by the different photometric bands varies with stellar population age.}
\label{fig:hst_filters}
\end{figure}

For clarity, Figs.~\ref{fig:spectra_lines} and \ref{fig:hst_filters} illustrate the location of the spectral segments and the photometric transmission bands used in the ablation analysis. These figures are intended to facilitate the interpretation of the results presented in Sect.~\ref{sec:Input dependencies} by showing the correspondence between the perturbed wavelength regions, the main spectral features, and the HST photometric filters.

\section{Ablation test effect on the mass-weighted ages }\label{app:ablation_metallicity}
\begin{figure*}
    \centering
    \includegraphics[width=0.99\linewidth]{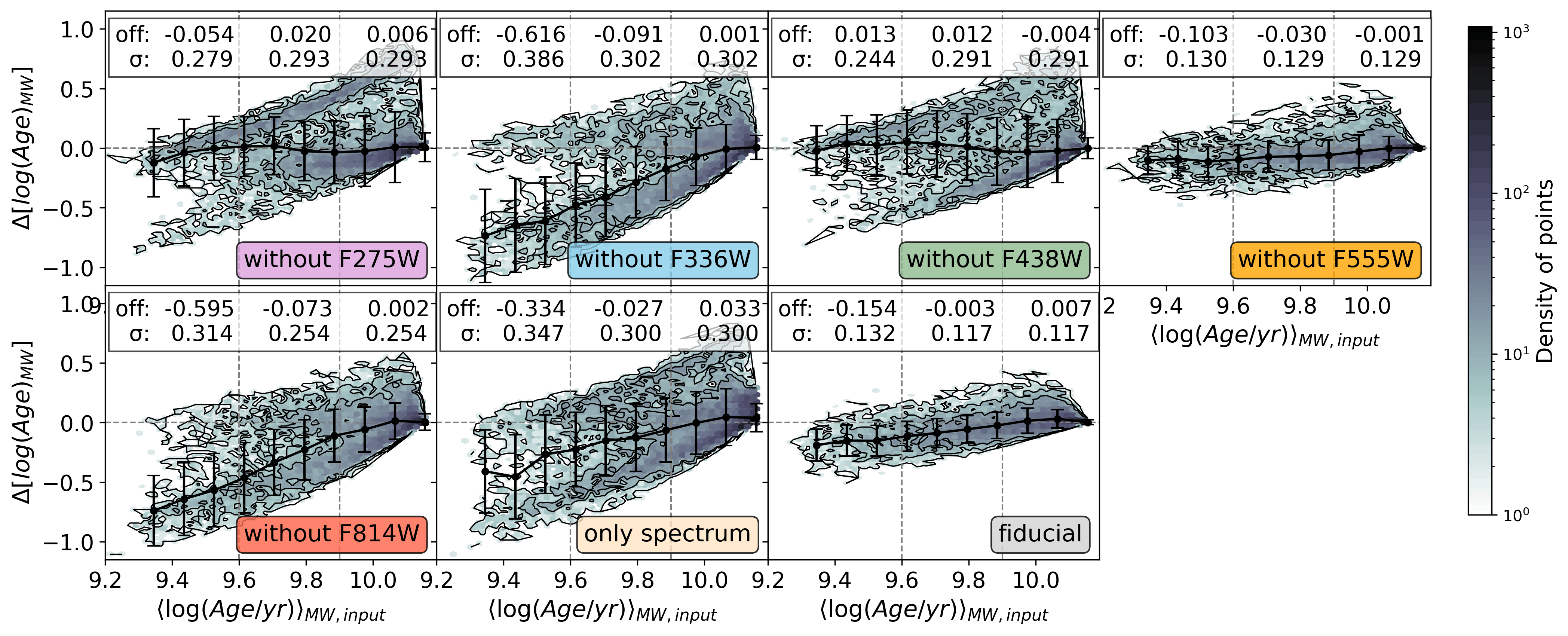}
    \caption{Impact of removing photometric fluxes from the input data on the recovery of the mass-weighted age. Each panel shows the difference between the predicted and input mass-weighted age as a function of the input value, when individual photometric fluxes are replaced by values randomly drawn from the empirical distribution of the whole sample. Additionally, the full photometric input can be replaced with random values, leaving only the spectrum unchanged. For reference, the fiducial case in which all photometric fluxes are included is also shown. The running median and standard deviation are plotted as black circles with error bars. The insets report the mean bias and standard deviation in three age intervals, whose boundaries are marked by vertical lines: (1) $\langle \log(\mathrm{Age}/\mathrm{yr}) \rangle_{\rm MW} \leq 9.6$; (2) $9.6 < \langle \log(\mathrm{Age}/\mathrm{yr}) \rangle_{\rm MW} \leq 9.9$; and (3)  $ \langle \log(\mathrm{Age}/\mathrm{yr}) \rangle_{\rm MW} > 9.9$.}
    \label{fig:ablation-age-mw-phot}
\end{figure*}

Figure \ref{fig:ablation-age-mw-phot}  shows the mean differences between the input and predicted ages when the photometric fluxes or sections of the spectra are substituted by other values, randomly taken from a different SFH of the training set. Figures contain the same information as Figs.~\ref{fig:ablation_age_lw_phot} and \ref{fig:violin_means_spectra_secciones} but for the MW version of the mean ages. 

\begin{figure*}
    \centering
   \includegraphics[width=17cm]{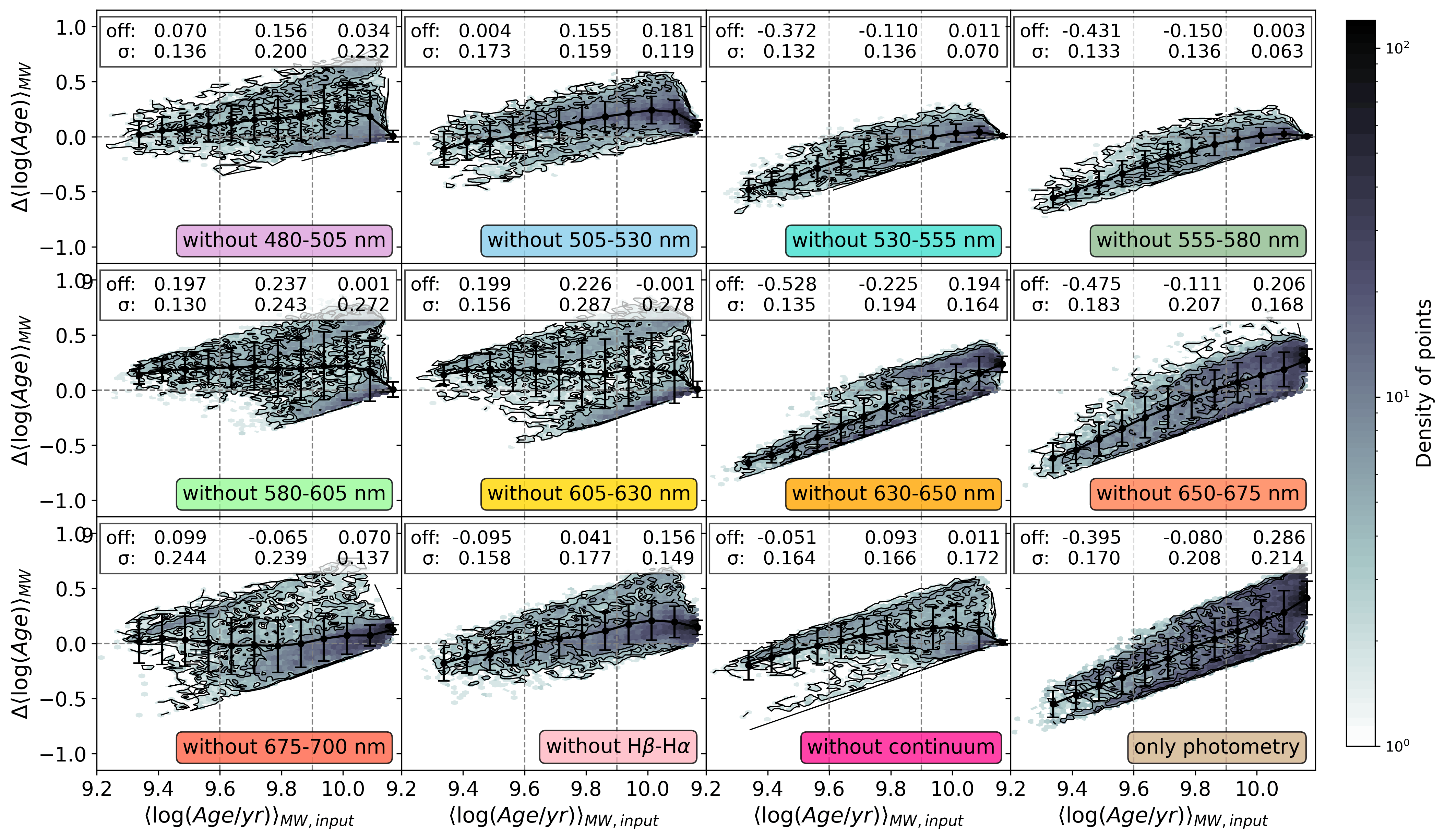}
    \caption{Impact of different spectral regions on the MW-age predictions. Each panel shows the difference between the ground-truth and predicted $\langle \log(\mathrm{Age}/\mathrm{yr}) \rangle_{\rm MW}$ for different perturbations of the input spectrum. The first nine panels (from left to right and top to bottom) show the results obtained when individual wavelength segments are replaced by random values drawn from the empirical distribution of the same segment across the full dataset. The panel labelled ``without continuum'' shows the results when the continuum shape is removed from the spectra. The final panel shows the case in which the full spectrum is replaced by random values. The running median and standard deviation are plotted as black circles with error bars. For each configuration, the mean offset and scatter are listed at the top of each panel for three age intervals, delineated by the vertical lines: (1) $\langle \log(\mathrm{Age}/\mathrm{yr}) \rangle_{\rm MW} \leq 9.6$; (2) $9.6 < \langle \log(\mathrm{Age}/\mathrm{yr}) \rangle_{\rm MW} \leq 9.9$; and (3)  $ \langle \log(\mathrm{Age}/\mathrm{yr}) \rangle_{\rm MW} > 9.9$.}
    \label{fig:ablation-age-mw-spec}
\end{figure*}

\begin{figure*}[ht]
    \centering
   \includegraphics[width=17cm]{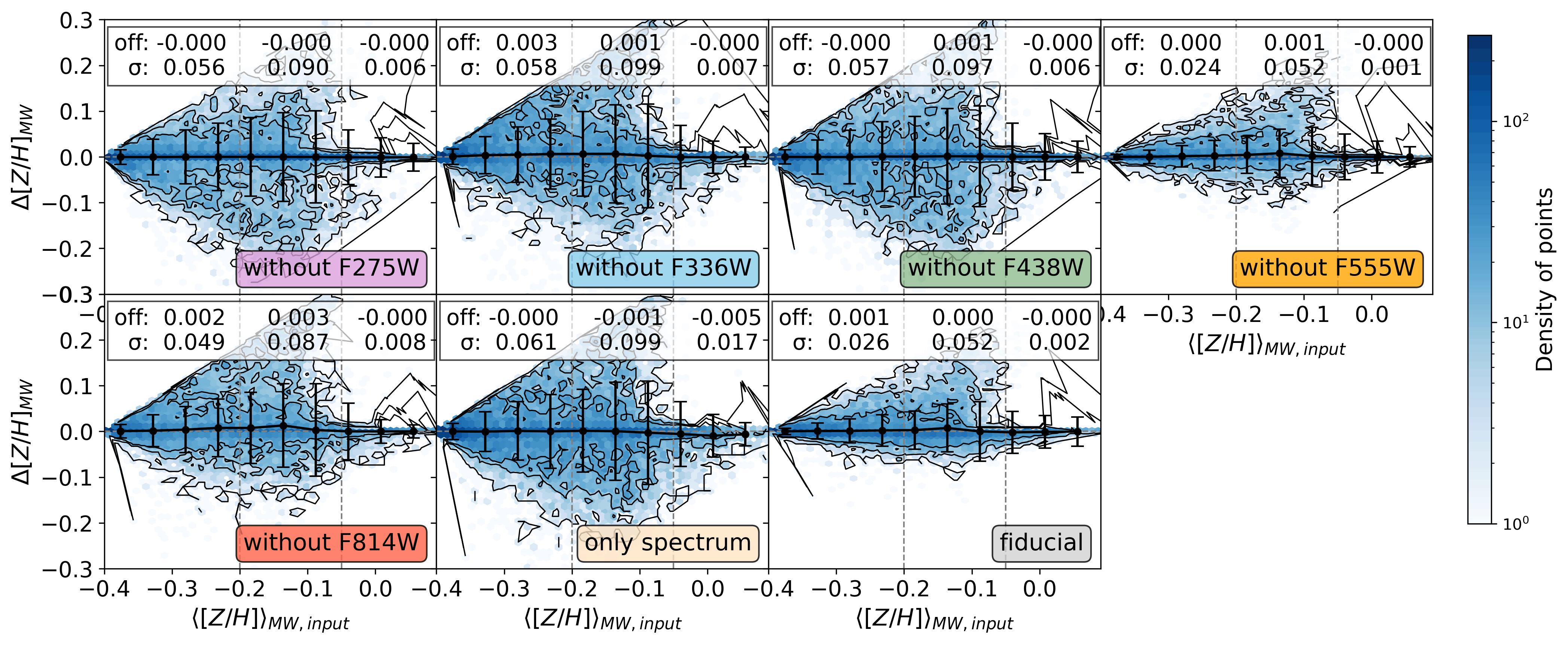}
 \caption{Same as Fig.~\ref{fig:ablation-age-mw-phot}, but for the mass-weighted metallicity. The insets report the mean bias and standard deviation in three metallicity intervals, whose boundaries are marked by vertical lines: (1) $\langle [Z/H] \rangle_{\rm LW} \leq -0.2$; (2)  $-0.2 < \langle [Z/H] \rangle_{\rm LW} \leq 0.05$; and (3)  $ \langle \langle [Z/H] \rangle_{\rm LW} > 0.05$.}
    \label{fig:ablation-z-phot}
\end{figure*}

\begin{figure*}[ht]
    \centering
    \includegraphics[width=17cm]{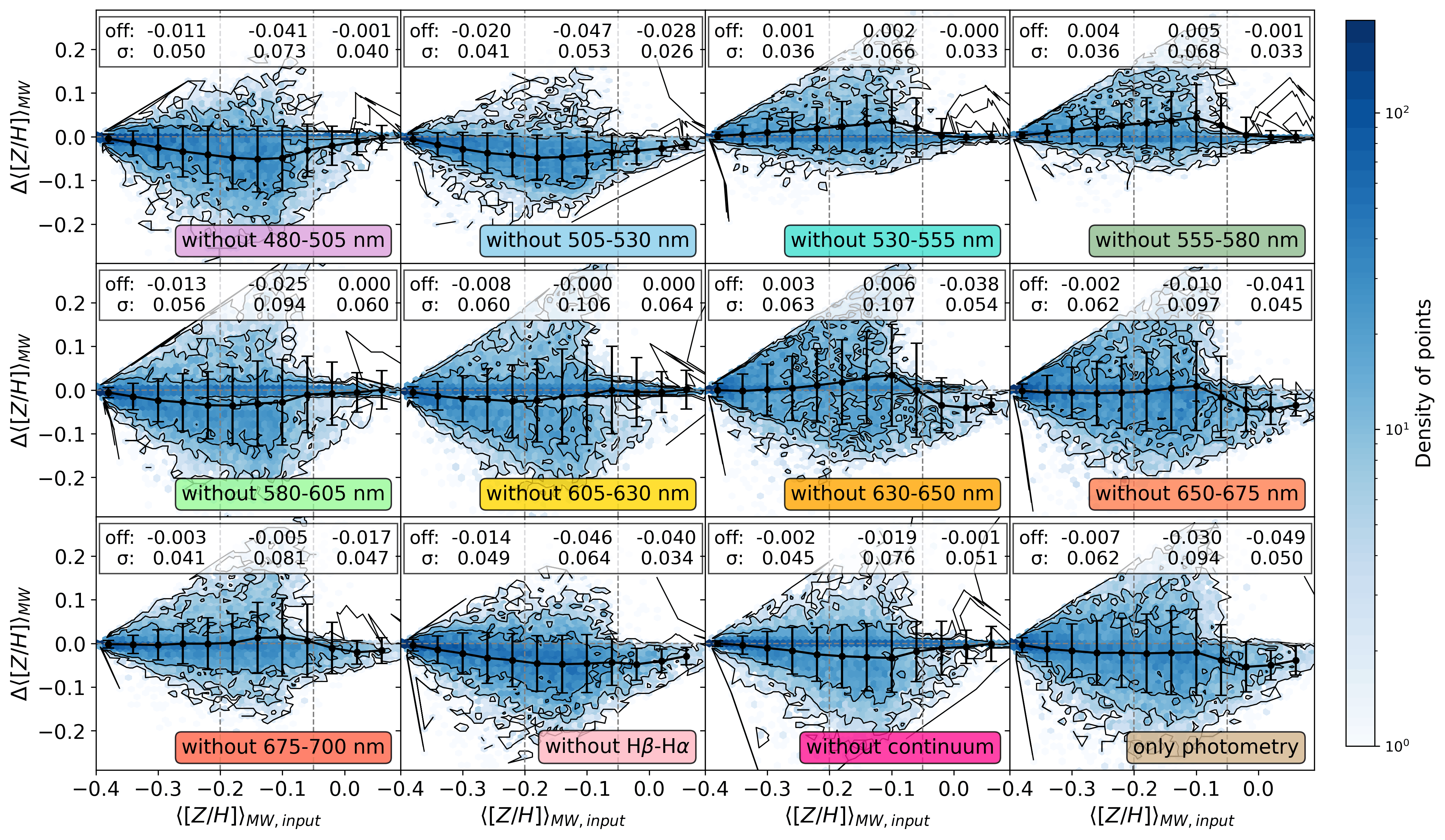}
    \caption{Impact of different spectral regions on the predictions of the mean MW metallicity. The different panels are equivalent to those in Fig.~\ref{fig:ablation-age-mw-spec}, but for $\langle [Z/H] \rangle_{\rm MW}$. Masking most wavelength intervals produces only minor changes in accuracy, while removing broad continuum information or relying on photometry alone leads to larger scatter, demonstrating the complementary role of spectral and photometric data in constraining metallicity. The insets report the mean bias and standard deviation in three metallicity intervals (see Fig.~\ref{fig:ablation-z-phot}).}
    \label{fig:ablation-z-spec}
\end{figure*}

Figs. \ref{fig:ablation-z-phot} and \ref{fig:ablation-z-spec} show the results of the ablation tests for the metallicity predictions. 

%
%



\end{document}